\documentclass[12pt]{article}
\usepackage{graphicx}
\usepackage[cp1251]{inputenc}
\usepackage{epsfig}
\usepackage[english]{babel}

\date{}
\def\b{\begin{equation}}
\def\e{\end{equation}}
\def\bee{\begin{enumerate}}
\def\eee{\end{enumerate}}
\def\be{\begin{vmatrix}}
\def\ee{\end{vmatrix}}

\begin{document}
\setcounter{page}{1}
\bibliographystyle{unsrt2}
\pagestyle{plain}

\title{\bf{Spin structure factors of doped
 monolayer Germanene in the presence of spin-orbit coupling}}
\author{Farshad Azizi and Hamed Rezania\thanks{Corresponding author.
Tel./fax: +98 831 427 4569., Tel: +98 831 427 4569}}
\maketitle{\centerline{Department of Physics, Razi University,
Kermanshah, Iran}
Corresponding author's email: rezania.hamed@gmail.com
\begin{abstract}
In this paper, we present a Kane-Mele model in the presence of
magnetic field and next nearest neighbors hopping amplitudes for
 investigations of the spin
 susceptibilities of Germanene layer.
Green's function approach has been implemented
to find the behavior of dynamical spin susceptibilities of Germanene layer
 within linear response theoryand in the presence of magnetic
 field and spin-orbit coupling
 at finite temperature.
Our results show the magnetic excitation mode for both longitudinal and transverse
 components of spin
tends to higher frequencies with spin-orbit coupling strength.
Moreover the frequency positions of sharp peaks in longitudinal dynamical spin susceptibility
are not affected by variation of magnetic field while the peaks in transverse
 dynamical susceptibility moves to lower frequencies with magnetic field.
The effects of electron doping on frequency behaviors of spin susceptibilities have been
addressed in details.
Finally the temperature dependence of static spin structure factors due to the
effects of spin-orbit coupling, magnetic field and chemical potential has been
studied.

\end{abstract}
\vspace{0.5cm} {\it \emph{Keywords}}: Germanene; Green's function; Optical absorption
\section{Introduction}
A lot of theoretical and experimental studies have been performed on
Graphene as a one-atom-thick layer of graphite since it's fabrication\cite{novoselov}.
The low energy linear dispersion and chiral property of carbon structure
leads to map the nearest neighbor hopping tight binding hamiltonian which
at low energy to a relativistic Dirac Hamiltonian for massless
 fermions with Fermi velocity $v_{F}$.
Novel electronic properties have been exhibited by
Graphene layer with a zero band gap which
compared to materials with a non-zero energy gap.
 These materials have intriguing physical properties and
numerous potential practical applications in optoelectronics and sensors\cite{wang}.

Recently, the hybrid systems consisting of Graphene and various two-dimensional
materials have been studied extensively both experimentally and theoretically
\cite{liu,chang1,chang2}.
Also, 2D materials could be used for a extensive applications in
nanotechnology \cite{dean,novo} and memory technology \cite{bertol}.
While the research interest in Graphene-based superlattices is growing rapidly,
 people have started
to question whether the Graphene could be replaced by its close relatives, such as 2 dimensional
hexagonal crystal of Germanene. This material shows a zero gap semiconductor with massless fermion
charge carriers since their $\pi$ and $\pi^{*}$ bands are also linear at
 the Fermi level\cite{akturk}.
Germanene as counterpart of Graphene, is predicted to have a geometry with low-buckled
honeycomb structure for its most stable structures in contrast to the Graphene monolayer
\cite{akturk,ccliu}. Such small buckling as vertical distance between two planes of
atoms for Germanene comes from the mixing of $sp^{2}$ and
$sp^{3}$ hybridization\cite{padil,chowd}. The behavior of Germanene electronic structure shows a linear
dispersion close to K and K' points of the first Brillouin zone. 
However {\it ab initio} calculations indicated
that spin-orbit coupling in Germanene causes to small band gap opening at the Dirac
point and thus the Germanene has massive Dirac fermions\cite{ccliu, kaloni}.
Also the band gap due to the spin orbit coupling in Germanene is more remarkable
 rather than that in Graphene\cite{liu1}.
The intrinsic carrier mobility of Germanene is higher than Graphene\cite{ye}.
The different dopants within the Germanene layer gives arise to the sizable band gap opening at the
Dirac point an the electronic properties of this material are affected by that\cite{monshi,sun}.
In a theoretical work, the structural and electronic properties of superlattices made
with alternate stacking of Germanene layer are systematically investigated by using a density
functional theory with the van der Waals correction\cite{xia}.
 It was predicted that spin orbit coupling and exchange field together open a
nontrivial bulk gap in Graphene like structures leading to the quantum spin hall effect
\cite{qiao,tse}. The topological phase transitions in the 2D crystals can be understood
based on intrinsic spin orbit coupling which arises due to
 perpendicular electric field or interaction with a substrate.
Kane and Mele\cite{kane} applied a model Hamiltonian to describe topological insulators.
Such model consists of a hopping and an intrinsic spin-orbit term on the Graphene like structures.
The Kane-Mele model essentially includes two copies with different sign for up and down spins of
a model introduced earlier by Haldane\cite{haldan}.
Such microscopic model was originally proposed
to describe the quantum spin Hall effect in Graphene\cite{kane}. Subsequent band structure
calculations showed, however, that the spin orbit gap in Graphene is so small\cite{min,yao} that the quantum spin
Hall effect in Graphene like structures is beyond experimental relevance.

In-plane magnetic field affects the magneto conductivity of honeycomb structures so that 
the results show the negative for intrinsic gapless Graphene. However the
 magneto-resistance of gapless Graphene presents a positive value for fields lower than the critical magnetic
field and negative above the critical magnetic\cite{hwang}.
Moreover, microwave magneto transport in doped Graphene
is an open problem\cite{mani}.

The many body effects such as Coulomb interaction and its dynamical screening
present the novel features for any electronic material.
The collective spectrum and quasiparticle properties of electronic systems are determined form dynamical
spin structure factors. These are applied to imply the optical
properties of the system. Moreover a lot of studies have
been done on collective modes of monolayer Graphene both
theoretically \cite{hwang22} and experimentally\cite{wilis}. However there are no
 the extensive theoretical studies
  on doped bilayer systems.

The frequency dependence of dynamical spin susceptibility
has been studied and the results causes to
 find the collective magnetic excitation spectrum of
many body system.
It is worthwhile to explain the experimental
interpretation of imaginary part of dynamical spin susceptibilities.
Slow neutrons scatter from solids via magnetic dipole interaction in
which the magnetic moment of the neutron interacts with the spin
magnetic moment of electrons in the solid\cite{doniach}. We can readily express the
inelastic cross-section of scattering of neutron beam from a
magnetic system based on
correlation functions between spin density operators.
In other words
the differential inelastic cross
section $d^{2}\sigma/d\Omega d\omega$ corresponds to imaginary
part of spin susceptibilities. $\omega$ describes the energy loss of
neutron beam which is defined as the difference between incident and
scattered neutron energies. $\Omega$ introduces solid angle of
scattered neutrons. The spin excitation
modes of the magnetic system have been found via the frequency position of peaks in
$d^{2}\sigma/d\Omega d\omega$. Depending on component of spin
magnetic moment of electrons that interacts with spin of neutrons,
transverse and longitudinal spin susceptibility behaviors have been
investigated.
The imaginary and real part of non-interacting
change susceptibilities of Graphene within an analytical
approach have been calculated
 A theoretical work
has been performed for calculating both \cite{peres1}.
The results of this study shows there is no remarkable
angle dependence for imaginary part of polarizability around the van
Hove singularity, i.e, $\hbar\omega/t=2.0$ where $t$ implies nearest
neighbor hopping integral.

It is worthwhile to add few comments regarding the comparison Germanene and Graphene like structure.
As we have mentioned, the most important difference between Germanene structure ad Graphene one arises from the nonzero
 overlap function between nearest neighbor atoms in Germanene structure.
Response functions of Graphene in the presence of spin-orbit coupling have been studied recently\cite{inglot,inglot1}.
In this references, the optical absorption of Graphene structure in the presence of spin-orbit coupling and magnetic field has been
theoretically studied. Optical absorption rate corresponds to the charge transition rate of electrons between energy levels.
 However in our work, we have investigated the dynamical spin susceptibility of Germanene structure due to spin-orbit coupling.
In other words we have specially studied the transition rate of magnetic degrees freedom of electrons.
Such study is a novelty of our work so that there is no the theoretical work on
 the study of magnetic excitation modes in Germanene structure due to spin-orbit coupling. These studies have been performed for Graphene like structures and
the most important difference between our results and the spin susceptibilities of Graphene structure is the effects of
overlap function in Germanene structure compared to Graphene lattice. In fact overlap function has considerable impact on the frequency position
of excitation mode and also on the intensity of scattered neutron beam from Germanene structure.

The purpose of this paper is to provide a
 Kane Mele model including intrinsic spin-orbit interaction
for studying frequency behavior of dynamical spin susceptibility of Germanene layer
 in the presence of
magnetic field perpendicular to the plane.
Using the suitable hopping integral and on site parameter values, the band dispersion of
electrons has been calculated. Full band calculation beyond Dirac approximation has
been implemented to derive both transverse and longitudinal
 dynamical spin susceptibilities. We have exploited Green's function approach to calculate the
spin susceptibility, i.e. the time ordered spin operator correlation.
 The effects of electron doping, magnetic field and spin-orbit coupling on the spin structure factors
 have been studied. Also we discuss and analyze to
show how spin-orbit coupling affects the frequency behavior of the
longitudinal and transverse spin susceptibilities. Also we study
 the frequency behavior of dynamical spin susceptibility
of Germanene due to variation of
chemical potential and magnetic field.
Also the effects of spin-orbit coupling constant
and magnetic field on temperature dependence of both transverse and longitudinal
static spin susceptibilities have been investigated in details.
 \section{Model Hamiltonian and formalism}
The crystal structure of Germanene has been shown in Fig.(1).
The unit cell of Germanene structure is similar to Graphene layer and this honeycomb lattice
 depicted in Fig.(2). The primitive unit cell vectors of honeycomb lattice
have been shown by ${\bf a}_{1}$ and ${\bf a}_{2}$.
In the presence of longitudinal magnetic field,
the Kane-Mele model\cite{kane} ($H$)
for Germanene structure includes the tight binding model ($H^{TB}$), the intrinsic spin-orbit coupling ($H^{ISOC}$)
and the Zeeman term ($H^{Zeeman}$) due to the coupling of spin degrees of freedom of electrons with external longitudinal magnetic field
 $B$
\begin{eqnarray}
 H&=&H^{TB}+H^{ISOC}+H^{Zeeman}.
 \label{e0.5}
\end{eqnarray}
The tight binding part of model Hamiltonian
consists of three parts; nearest
neighbor hopping, next nearest neighbor (2NN) hopping and
 next next nearest neighbor (3NN) hopping terms. The tight binding part, the spin orbit coupling term
and the Zeeman part of the model Hamiltonian on the honeycomb lattice are given by
\begin{eqnarray}
 H^{TB}&=&-t\sum_{i,{\bf\Delta},\sigma}\Big(a^{\sigma\dag}_{j=i+{\bf\Delta}}b^{\sigma}_{i}+h.c.\Big)
-t'\sum_{i,{\bf\Delta}',\sigma}\Big(a^{\sigma\dag}_{j=i+{\bf\Delta}'}a^{\sigma}_{i}+b^{\sigma\dag}_{i+{\bf\Delta}'}b^{\sigma}_{i}\Big)
-t"\sum_{i,{\bf\Delta}",\sigma}\Big(a^{\sigma\dag}_{j=i+{\bf\Delta}"}b^{\sigma}_{i}+h.c.\Big)
\nonumber\\&-&\sum_{i,\sigma}
\mu\Big(a^{\dag\sigma}_{i}a^{\sigma}_{i}+b^{\dag\sigma}_{i}
b^{\sigma}_{i}\Big),\nonumber\\
H^{ISOC}&=&
i\lambda\sum_{i,{\bf\Delta}',\sigma}\sum_{\alpha=A,B}\Big(\nu^{a}_{i+\Delta',i}
a^{\dag\sigma}_{i+{\bf\Delta}'}\sigma^{z}_{\sigma\sigma'}
a^{\sigma'}_{i}+\nu^{b}_{i+{\bf\Delta}',i}
b^{\dag\sigma}_{i+{\bf\Delta}'}\sigma^{z}_{\sigma\sigma'}
b^{\sigma'}_{i}\Big),\nonumber\\
H^{Zeeman}&=&-\sum_{i,\sigma}\sigma g\mu_{B}B\Big(a^{\dag\sigma}_{i}a^{\sigma}_{i}+b^{\dag\sigma}_{i}
b^{\sigma}_{i}\Big).
\label{e1}
\end{eqnarray}
Here $a^{\sigma}_{i}(b^{\sigma}_{i})$ is an annihilation operator of electron with
spin $\sigma$ on sublattice
$A(B)$ in unit cell index $i$. The operators fulfill the fermionic standard
anti commutation relations $\{a^{\sigma}_{i},a^{\sigma'\dag}_{j}\}=
\delta_{ij}\delta_{\sigma\sigma'}$. As usual $t,t',t''$ denote the nearest neighbor, next
nearest neighbor and next next nearest neighbor
 hopping integral amplitudes, respectively. The parameter $\lambda$ introduces the
spin-orbit coupling strength. Also $B$ refers to strength of applied magnetic field.
$g$ and $\mu_{B}$ introduce the gyromagnetic and Bohr magneton constants, respectively.
 $\sigma^{z}$ is the third Pauli matrix, and $\nu_{ji}^{a(b)}=\pm 1$
as discussed below.
Based on Fig.(2), ${\bf a}_{1}$ and ${\bf a}_{2}$ are the primitive vectors of unit
cell and the length of them is assumed to be unit.
The symbol ${\bf \Delta}={\bf 0},{\bf\Delta}_{1},{\bf\Delta}_{2}$
 implies the indexes of
lattice vectors connecting the unit cells including nearest neighbor lattice sites.
The translational vectors  ${\bf\Delta}_{1},{\bf\Delta}_{2}$ connecting
 neighbor unit cells are given by
\begin{eqnarray}
{\bf\Delta}_{1}={\bf i}\frac{\sqrt{3}}{2}+{\bf j}\frac{1}{2}\;\;,\;\;{\bf\Delta}_{2}=
{\bf i}\frac{\sqrt{3}}{2}-{\bf j}\frac{1}{2}.
\label{e4}
\end{eqnarray}
Also index ${\bf\Delta}'={\bf\Delta}_{1},{\bf\Delta}_{2},
-{\bf\Delta}_{1},-{\bf\Delta}_{2},{\bf j},-{\bf j}$
implies the characters of lattice
vectors connecting the unit cells including next nearest neighbor lattice sites.
Moreover index ${\bf\Delta}"=\sqrt{3}{\bf i},{\bf j},-{\bf j}$
denotes the characters of lattice
vectors connecting the unit cells including next next nearest neighbor lattice sites.
We consider the intrinsic spin-orbit term\cite{kane}
 of the KM Hamiltonian in Eq.(\ref{e1}). The expression $\nu_{ji}^{a(b)}$ gives $\pm 1$
depending on the orientation of the sites. A standard definition for $\nu^{\alpha}_{ji}$ in each sublattice $\alpha=A,B$ is $\nu_{ji}^{\alpha}=\Big(
\frac{{\bf d}^{\alpha}_{j}\times {\bf d}^{\alpha}
_{i}}{|{\bf d}^{\alpha}_{j}\times {\bf d}^{\alpha}_{i}|}\Big).{\bf e}_{z}=\pm 1$ where ${\bf d}
^{\alpha}_{j}$ and ${\bf d}
^{\alpha}_{i}$ are the two unit vectors along the nearest neighbor bonds connecting site $i$
to its next-nearest neighbor $j$. Moreover ${\bf e}_{z}$ implies the unit vector perpendicular to the plane.
  Because of two sublattice atoms, the band wave function
$\psi^{\sigma}_{n}({\bf k},{\bf r})$ can be expanded in terms of Bloch functions
$\Phi^{\sigma}_{\alpha}({\bf k},{\bf r})$. The index $\alpha$ implies two inequivalent sublattice atoms
$A,B$ in the unit cell, ${\bf r}$ denotes the position vector of electron, ${\bf k}$ is
the wave function belonging in the first Brillouin zone of honeycomb structure.
Such band wave function can be written as
\begin{eqnarray}
\psi^{\sigma}_{n}({\bf k},{\bf r})=\sum_{\alpha=A,B}C^{\sigma}_{n\alpha}({\bf k})
\Phi^{\sigma}_{\alpha}({\bf k},{\bf r}),
 \label{e5}
\end{eqnarray}
where $C^{\sigma}_{n\alpha}({\bf k})$ is the expansion coefficients and $n=c,v$ refers
 to condition
and valence bands. Also we expand the Bloch wave function in terms of Wannier wave function as
\begin{eqnarray}
\Phi^{\sigma}_{\alpha}({\bf k},{\bf r})=\frac{1}{\sqrt{N}}\sum_{{\bf R}_{i}}
e^{i{\bf k}.{\bf R}_{i}}\phi^{\sigma}_{\alpha}({\bf r}-{\bf R}_{i}),
 \label{e6}
\end{eqnarray}
so that ${\bf R}_{i}$ implies the position vector of $i$th unit cell in the crystal and
$\phi_{\alpha}$ is the Wannier wave function of electron in the vicinity of
atom in $i$ th unit cell on sublattice index $\alpha$.
By inverting the expansion Eq.(\ref{e5}), we can expand the Bloch wave functions
 in terms of band wave function as following
relation
\begin{eqnarray}
\Phi^{\sigma}_{\alpha}({\bf k},{\bf r})=\sum_{n=c,v}D_{\alpha n}^{\sigma}({\bf k})
\psi^{\sigma}_{n}({\bf k},{\bf r}),
\label{e6.5}
\end{eqnarray}
where $D_{\alpha n}^{\sigma}({\bf k})$
 is the expansion coefficients and we explain these coefficients in the following.

 The small Buckling in Germanene causes
to the considerable value for 2NN and 3NN hopping amplitude. Moreover we have considerable values for overlap parameters of electron
wave functions between 2NN and 3NN atoms. The band structures of electrons with spin $\sigma$
 of Germanene described by model Hamiltonian in Eq.(\ref{e1}) are
obtained by using the matrix form of Schrodinger as follows
\begin{eqnarray}
{\mathcal H}^{\sigma}({\bf k}){\mathcal C}^{\sigma}({\bf k})&=&E^{\sigma}_{n}({\bf k}){\mathcal S}^{\sigma}
({\bf k}){\mathcal C}^{\sigma}({\bf k}),\nonumber\\
{\mathcal H}^{\sigma}({\bf k})&=&\left(
                              \begin{array}{cc}
                               H^{\sigma}_{AA}({\bf k})&   H^{\sigma}_{AB}({\bf k}) \\
                               H_{BA}^{\sigma}({\bf k}) &  H^{\sigma}_{BB}({\bf k}) \\
\end{array}
\right)\;,\;{\mathcal C}^{\sigma}({\bf k})=\left(
                              \begin{array}{c}
                               C_{nA}^{\sigma}({\bf k})\\
                               C_{nB}^{\sigma}({\bf k})\\
\end{array}
\right),\;\;\nonumber\\
{\mathcal S}^{\sigma}({\bf k})&=&\left(
                              \begin{array}{cc}
                               S^{\sigma}_{AA}({\bf k})&   S^{\sigma}_{AB}({\bf k}) \\
                               S^{\sigma}_{BA}({\bf k}) &  S^{\sigma}_{BB}({\bf k})\\
\end{array}
\right).
 \label{e7}
\end{eqnarray}
Using the Bloch wave functions, i.e. $\Phi_{\alpha}({\bf k})$, the matrix elements of ${\mathcal H}$ and ${\mathcal S}$ are given by
\begin{eqnarray}
 H^{\sigma}_{\alpha\beta}({\bf k})=\langle\Phi^{\sigma}_{\alpha}({\bf k})|H|\Phi^{\sigma}_{\beta}({\bf k})\rangle\;\;,\;\;
 S^{\sigma}_{\alpha\beta}({\bf k})=\langle\Phi^{\sigma}_{\alpha}({\bf k})|\Phi^{\sigma}_{\beta}({\bf k})\rangle.
 \label{e8}
\end{eqnarray}
The matrix elements of $H^{\sigma}_{\alpha\beta}$ and $S^{\sigma}_{\alpha\beta}$ are expressed based on hopping amplitude and spin-orbit coupling
between two neighbor atoms on lattice sites and can be expanded in terms of hopping amplitudes $t,t',t"$,
spin orbit coupling $\lambda$ and overlap parameters. The diagonal elements of matrixes ${\mathcal H}$
 in Eq.(\ref{e7}) arise from hopping amplitude of electrons between next nearest neighbor atoms on the same sublattice
 and spin-orbit coupling.
Also the off diagonal matrix elements with spin channel $\sigma$, i.e. $H^{\sigma}_{AB}, H^{\sigma}_{BA}$, raise from hopping amplitude of electrons between
 nearest neighbor atoms and next next nearest neighbor atoms on the different sublattices.
These matrix elements are obtained as
\begin{eqnarray}
H^{\sigma}_{AB}({\bf k})&=&t\Big(1+e^{i{\bf k}.{\bf\Delta}_{1}}+e^{i{\bf k}.{\bf\Delta}_{2}}\Big)+t^{"}
\Big(2\cos(k_{y})+e^{-i\sqrt{3}k_{x}}\Big)\nonumber\\
&=&t\Big(1+2\cos(k_{y}/2)e^{-i\sqrt{3}k_{x}/2}\Big)+t^{"}
\Big(2\cos(k_{y})+e^{-i\sqrt{3}k_{x}}\Big),\nonumber\\
H^{\sigma}_{AA}({\bf k})&=&2t'\Big(\cos(\sqrt{3}k_{x}/2+k_{y}/2)+\cos(\sqrt{3}k_{x}/2+k_{y}/2)+\cos(k_{y}/2)\Big)\nonumber\\
&-&2\lambda \Big(\sin\Big(\frac{1}{2}k_{y}\Big)-
\sin\Big(\frac{\sqrt{3}}{2}k_{x}+\frac{1}{2}k_{y}\Big)-
\sin\Big(\frac{\sqrt{3}}{2}k_{x}-\frac{1}{2}k_{y}\Big)\Big)-\mu-\sigma g\mu_{B}B,\nonumber\\
H^{\sigma}_{BB}({\bf k})&=&-2t'\Big(\cos(\sqrt{3}k_{x}/2+k_{y}/2)+\cos(\sqrt{3}k_{x}/2+k_{y}/2)+\cos(k_{y}/2)\Big)\nonumber\\
&+&2\lambda \Big(\sin\Big(\frac{1}{2}k_{y}\Big)-
\sin\Big(\frac{\sqrt{3}}{2}k_{x}+\frac{1}{2}k_{y}\Big)-
\sin\Big(\frac{\sqrt{3}}{2}k_{x}-\frac{1}{2}k_{y}\Big)\Big)-\mu-\sigma g\mu_{B}B,\nonumber\\
H^{\sigma}_{BA}({\bf k})&=&H^{*}_{AB}({\bf k}).
 \label{e9}
\end{eqnarray}
Based on matrix elements $H^{\sigma}_{\alpha\beta}({\bf k})$, the model Hamiltonian in Eq.(\ref{e1}) is
written in terms of Fourier transformation of creation and annihilation fermionic operators as
\begin{eqnarray}
H=\sum_{{\bf k},\sigma}\Big[H^{\sigma}_{AA}({\bf k})a^{\sigma\dag}_{\bf k}a^{\sigma}_{\bf k}
+H_{AB}({\bf k})a^{\sigma\dag}_{\bf k}b^{\sigma}_{\bf k}+
H_{BA}({\bf k})b^{\sigma\dag}_{\bf k}a^{\sigma}_{\bf k}+
H^{\sigma}_{BB}({\bf k})b^{\sigma\dag}_{\bf k}b^{\sigma}_{\bf k}\Big],
 \label{e9.1}
\end{eqnarray}
so that the operator $a^{\sigma}_{\bf k}(b^{\sigma}_{\bf k})$ annihilates an electron at wave vector
${\bf k}$ with spin index $\sigma$ on sublattice A(B) and has the following relation as
\begin{eqnarray}
a^{\sigma}_{\bf k}=\frac{1}{\sqrt{N}}\sum_{\bf k}e^{i{\bf k}\cdot{\bf R}_{i}}a^{\sigma}_{i}
\;\;,\;\;b^{\sigma}_{\bf k}=\frac{1}{\sqrt{N}}\sum_{\bf k}e^{i{\bf k}\cdot{\bf R}_{i}}b^{\sigma}_{i}.
 \label{e9.2}
\end{eqnarray}

The matrix elements of ${\mathcal S}({\bf k})$, i.e. $S_{AA}({\bf k})$
, $S_{AB}({\bf k})$, $S_{BA}({\bf k})$
and $S_{BB}({\bf k})$ are expressed as
\begin{eqnarray}
S_{AB}({\bf k})&=&s\Big(1+e^{i{\bf k}.{\bf\Delta}_{1}}+e^{i{\bf k}.{\bf\Delta}_{2}}\Big)+s^{"}
\Big(2\cos(k_{y})+e^{-i\sqrt{3}k_{x}}\Big)\nonumber\\
&=&s\Big(1+2\cos(k_{y}/2)e^{-i\sqrt{3}k_{x}/2}\Big)+s^{"}
\Big(2\cos(k_{y})+e^{-i\sqrt{3}k_{x}}\Big)\nonumber\\
S_{AA}({\bf k})&=&1+2s'\Big(\cos(\sqrt{3}k_{x}/2+k_{y}/2)+\cos(\sqrt{3}k_{x}/2+k_{y}/2)+\cos(k_{y}/2)\Big)\nonumber\\
S_{BB}({\bf k})&=&S_{AA}({\bf k})\;\;,\;\;S_{BA}({\bf k})=S^{*}_{AB}({\bf k}),
 \label{e10}
\end{eqnarray}
so that $s$ is the overlap between orbital wave function of electron respect to the nearest neighbor atoms,
$s'$ denotes the overlap between orbital wave function of electron respect to the next nearest neighbor atoms and
$s"$ implies the overlap between orbital wave function of electron respect to the next next nearest neighbor atoms.
The density functional theory and {\it ab initio} calculations has been determined the hopping amplitudes and overlap values $s,s',s"$
as\cite{xia}$t=-1.163,t'=-0.055,t"=-0.0836,s=0.01207,s'=0.0128,s"=0.048$.
Using the
Hamiltonian and overlap matrix forms
 in Eqs.(\ref{e9},\ref{e10}), the band structure of electrons, i.e. $E^{\sigma}_{\eta}({\bf k})$
has been found by solving equation
 ${\rm\det}\Big({\mathcal H}({\bf k})-E({\bf k}){\mathcal S}({\bf k})\Big)=0$.
Moreover the matrix elements of $\mathcal{C}({\bf k})$ can be found based on eigenvalue equation
in Eq.(\ref{e7}).
Eq.(\ref{e7}) can be rewritten as matrix equation as follows
\begin{eqnarray}
 \left(
                              \begin{array}{cc}
                               \psi_{c}({\bf k},{\bf r}) \\
                                \psi_{v}({\bf k},{\bf r}) \\
\end{array}
\right)=\left(
                              \begin{array}{cc}
                              C^{\sigma}_{cA}({\bf k})&  C^{\sigma}_{cB}({\bf k}) \\
                               C^{\sigma}_{vA}({\bf k}) &  C^{\sigma}_{vB}({\bf k})\\
\end{array}
\right)\left(
                              \begin{array}{c}
                               \Phi^{\sigma}_{A}({\bf k},{\bf r})\\
                               \Phi^{\sigma}_{B}({\bf k},{\bf r})\\
\end{array}
\right).
\label{e10.2}
\end{eqnarray}
In a similar way, we can express the matrix from for Eq.(\ref{e6.5})
\begin{eqnarray}
 \left(
                              \begin{array}{cc}
                              \Phi^{\sigma}_{A}({\bf k},{\bf r})  \\
                                \Phi^{\sigma}_{B}({\bf k},{\bf r}) \\
\end{array}
\right)=\left(
                              \begin{array}{cc}
                              D^{\sigma}_{Ac}({\bf k})&  D^{\sigma}_{Av}({\bf k}) \\
                               D^{\sigma}_{Bc}({\bf k}) &  D^{\sigma}_{Bv}({\bf k})\\
\end{array}
\right)\left(
                              \begin{array}{c}
                               \psi^{\sigma}_{c}({\bf k},{\bf r})\\
                               \psi^{\sigma}_{v}({\bf k},{\bf r})\\
\end{array}
\right),\nonumber\\
\left(
                              \begin{array}{cc}
                              D^{\sigma}_{Ac}({\bf k})&  D^{\sigma}_{Av}({\bf k}) \\
                               D^{\sigma}_{Bc}({\bf k}) &  D^{\sigma}_{Bv}({\bf k})\\
\end{array}
\right)=\left(
                              \begin{array}{cc}
                              C^{\sigma}_{cA}({\bf k})&  C^{\sigma}_{cB}({\bf k}) \\
                               C^{\sigma}_{vA}({\bf k}) &  C^{\sigma}_{vB}({\bf k})\\
\end{array}
\right)^{-1}
\label{e10.3}
\end{eqnarray}
 The final results for band structure and expansion coefficients, i.e. $C^{\sigma}_{n\alpha}$
 and
$D^{\sigma}_{\alpha n}$,
are lengthy and are not given here. The valence and condition bands of electrons
 have been presented by $E^{\sigma}_{v}({\bf k})$ and $E^{\sigma}_{c}({\bf k})$ respectively.
In the second quantization representation, we can rewrite the Eq.(14) as
\begin{eqnarray}
 \left(
                              \begin{array}{cc}
                              a^{\sigma\dag}_{{\bf k}}  \\
                                b^{\sigma\dag}_{{\bf k}} \\
\end{array}
\right)=\left(
                              \begin{array}{cc}
                              D^{\sigma}_{Ac}({\bf k})&  D^{\sigma}_{Av}({\bf k}) \\
                               D^{\sigma}_{Bc}({\bf k}) &  D^{\sigma}_{Bv}({\bf k})\\
\end{array}
\right)\left(
                              \begin{array}{c}
                              c^{\sigma\dag}_{c,{\bf k}}\\
                               c^{\sigma\dag}_{v,{\bf k}}\\
\end{array}
\right),
\label{e10.4}
\end{eqnarray}

Using band energy spectrum,
the Hamiltonian in Eq.(\ref{e1}) can be rewritten by
\begin{eqnarray}
 H=\sum_{{\bf k},\sigma,\eta=c,v}
E^{\sigma}_{\eta}({\bf k})c^{\sigma\dag}_{\eta,{\bf k}}c^{\sigma}_{\eta,{\bf k}},
\label{e0.57}
\end{eqnarray}
where $c^{\sigma}_{\eta,{\bf k}}$ defines the creation operator of electron with spin
$\sigma$ in band index $\eta$ at wave vector ${\bf k}$.
Since longitudinal magnetic field has been applied perpendicular
 to the Germanene layer, the electronic Green's function depends on the
spin index $\sigma=\uparrow,\downarrow$.
According to the model Hamiltonian introduced in Eq.(\ref{e1}), the elements of
spin resolved Matsubara Green's function are introduced as the following forms
\begin{eqnarray}
G^{\sigma}_{AA}({\bf k},\tau)&=&-\langle {\cal T}(a_{{\bf k},\sigma}(\tau)a^{\dag}_{{\bf k},\sigma}(0))\rangle\;\;,\;\;
G^{\sigma}_{AB}({\bf k},\tau)=-\langle {\cal T}(a_{{\bf k},\sigma}(\tau)b^{\dag}_{{\bf k},\sigma}(0))\rangle,\nonumber\\
G^{\sigma}_{BA}({\bf k},\tau)&=&-\langle {\cal T}(b_{{\bf k},\sigma}(\tau)a^{\dag}_{{\bf k},\sigma}(0))\rangle\;\;,\;\;
G^{\sigma}_{BB}({\bf k},\tau)=-\langle {\cal T}(b_{{\bf k},\sigma}(\tau)b^{\dag}_{{\bf k},\sigma}(0))\rangle.
 \label{e3.3}
\end{eqnarray}
$T$ introduces the time ordering operator and arranges the creation and annihilation operators in terms of time of them without attention to the
their algebra.
The Fourier transformation of each Green's function element is obtained by
\begin{eqnarray}
 G^{\sigma}_{\alpha\beta}({\bf k},i\omega_{m})=\int^{1/k_{B}T}_{0}
d\tau e^{i\omega_{m}\tau}G^{\sigma}_{\alpha\beta}({\bf k},\tau)\;\;,\;\;\alpha,\beta=A,B,
\end{eqnarray}
 where $\omega_{m}=(2m+1)\pi k_{B}T$ denotes the Fermionic Matsubara frequency.
After some algebraic calculation, the following expression is obtained for Green's functions in Fourier presentation
\begin{eqnarray}
G^{\sigma}_{\alpha\beta}({\bf k},i\omega_{m})&=&\sum_{\eta=c,v}\frac{D_{\alpha\eta}^{\sigma *}({\bf k})
D^{\sigma}_{\beta\eta}({\bf k})}{i\omega_{m}-E^{\sigma}_{\eta}({\bf k})},
\label{e5.5}
\end{eqnarray}
where $\alpha,\beta$ refer to the each atomic basis of honeycomb lattice and $E^{\sigma}_{\eta}({\bf k})$ is the band structure of
Germanene layer in the presence of magnetic field and spin-orbit coupling.
For determining the chemical potential, $\mu_{\sigma}$,
we use the relation between concentration of electrons ($n_{e}$) and chemical potential.
This relation is given by
\begin{eqnarray}
 n_{e}=\frac{1}{4N}\sum_{{\bf k},\eta,\sigma}\frac{1}{e^{E^{\sigma}_{\eta}({\bf k})/k_{B}T}+1}.
\label{e0.63}
\end{eqnarray}
In fact the diagonal matrix elements of the Hamiltonian in Eq.(9) depends on chemical potential $\mu$. Thus eigenvalues, i.e.
$E^{\sigma}_{\eta}({\bf k})$ includes the factor $\mu$. Therefore the right hand of Eq.(20) depends on chemical potential $\mu$.
With an initial guess for chemical potential $\mu$, we can solve the algebraic equation 20 so that we can find the chemical potential value for
each amount for electronic concentration $n_{e}$. These statements have been added to the manuscript after Eq.(20).
Based on the values of
electronic concentration $n_{e}$, the chemical potential, $\mu$,
can be obtained by means Eq.(\ref{e0.63}).
In order to obtain the magnetic
excitation spectrum of Germanene structure both transverse and longitudinal
dynamical spin susceptibilities have been presented using Green's function method in the following section.
\section{Dynamical and static spin structure factors}
The correlation function between spin components of itinerant electrons in Germanene layer at different times
 can be expressed in terms of one particle Green's functions. The frequency Fourier transformation of this correlation function produces
 the dynamical spin susceptibility. The frequency position of peaks in dynamical spin susceptibility
 are associated with collective excitation of electronic gas described by Kane-Mele model Hamiltonian in the presence of magnetic field.
These excitations are related to the spin excitation
spectrum of electrons on Honeycomb structure.
 In the view point of experimental interpretation, the dynamical spin susceptibility of the localized electrons of the system
 is proportional to
 inelastic cross-section for magnetic
neutron scattering from a magnetic system that can be expressed in terms
of spin density correlation functions of the system. In other words the differential
inelastic cross section $d^{2}\sigma/d\Omega d\omega$ is
proportional to imaginary part of spin susceptibilities. $\omega$
denotes the energy loss of neutron beam which is defined as the
difference between incident and scattered neutron energies. The solid angle $\Omega$
implies the orientation of wave vector of scattered neutrons from the localized electrons of the sample.
We can assume the wave vector of incident neutrons is along $z$ direction.
The solid angle $\Omega$ depends on the polar angle between wave vector of scattered neutrons and the wave vector
of the incident neutrons.
 The frequency position
of peaks in $d^{2}\sigma/d\Omega d\omega$ determines the spin
excitation spectrum of the magnetic system\cite{doniach}.
In order to study the general spin excitation spectrum of the localized electron of the systems, both transverse and longitudinal
dynamical spin-spin correlation functions have been calculated.
 Linear response theory gives us the
 dynamical spin response functions based on the correlation function between
 components of spin operators. We introduce
$\chi_{+-}$ as transverse spin susceptibility and its relation is given by
\begin{eqnarray}
\chi_{+-}({\bf q},\omega)&=&i
\int_{-\infty}^{+\infty}dt e^{i\omega t}
\langle [S^{+}({\bf q},t),S^{-}(-{\bf q},0)]\rangle\nonumber\\&=&\lim_{i\Omega_{n}
\longrightarrow\omega+i0^{+}}\int_{0}^{1/(k_{B}T)}d\tau e^{i\Omega_{n}\tau}\langle
\mathcal{T}S^{+}({\bf q},\tau)S^{-}(-{\bf q},0)\rangle\nonumber\\&=&
\chi_{+-}({\bf q},i\Omega_{n}\longrightarrow\omega+i0^{+}),
 \label{e5.6}
\end{eqnarray}
in which $\Omega_{n}=2n\pi k_{B}T$ is the bosonic Matsubara frequency and
 $\mathcal{T}$ implies the time order operator.
Also the wave vector ${\bf q}$ in Eq.(\ref{e5.6}) implies the difference between incident and scattered neutron
wave vectors.
The Fourier transformations of transverse components
 of spin density operators, ($S^{+(-)}$), in terms of fermionic operators is given by
\begin{eqnarray}
S^{+}({\bf q})&=&\sum_{\bf k}\Big(a^{\dag\uparrow}_{{\bf k}+{\bf q}}
a^{\downarrow}_{\bf k}+b^{\dag\uparrow}_{{\bf k}+{\bf q}}
b^{\downarrow}_{\bf k}\Big)\;\;,\;\;
S^{-}({\bf q})=\sum_{\bf k}\Big(a^{\dag\downarrow}_{{\bf k}+{\bf q}}
a^{\uparrow}_{\bf k}+b^{\dag\downarrow}_{{\bf k}+{\bf q}}
b^{\uparrow}_{\bf k}\Big).
 \label{e5.7.1}
\end{eqnarray}
Substituting the operator form of $S^{+}$ and $S^{-}$ into
definition of transverse
spin susceptibility in Eq.(\ref{e5.6}), we arrive the following expression for transverse dynamical spin susceptibility
($\chi_{+-}({\bf q},i\Omega_{n})$)
\begin{eqnarray}
\chi_{+-}({\bf q},i\Omega_{n})&=&\int_{0}^{1/(k_{B}T)}d\tau e^{i\Omega_{n}\tau}\frac{1}{N^{2}}
\sum_{{\bf k},{\bf k}'}\Big\langle {\cal T}\Big(
a^{\dag\uparrow}_{{\bf k}+{\bf q}}(\tau)
a^{\downarrow}_{\bf k}(\tau)+b^{\dag\uparrow}_{{\bf k}+{\bf q}}(\tau)
b^{\downarrow}_{\bf k}(\tau)\Big)\nonumber\\&\times&
\Big(a^{\dag\downarrow}_{{\bf k}+{\bf q}}(0)
a^{\uparrow}_{\bf k}(0)+b^{\dag\downarrow}_{{\bf k}+{\bf q}}(0)
b^{\uparrow}_{\bf k}(0)\Big)
\Big\rangle,
\label{e0.701}
\end{eqnarray}
 that $N$ is the number of unit cells in Germanene structure.
In order to calculate the correlation function in Eq.(\ref{e0.701}),
one particle spin dependent Green's function matrix elements presented in Eq.(\ref{e5.5}) should be exploited.
After applying Wick's theorem and taking Fourier transformation, we can
transverse susceptibility in terms of one particle spin dependent
Green's function
\begin{eqnarray}
\chi_{+-}({\bf q},i\Omega_{n})=-
\frac{k_{B}T}{N}\sum_{{\bf k}}\sum_{\alpha,\beta}\sum_{m}
G^{\uparrow}_{\beta\alpha}({\bf k},i\omega_{m})G^{\downarrow}_
{\alpha\beta}({\bf k}+{\bf q},i\Omega_{n}+i\omega_{m}).
\label{e0.702}
\end{eqnarray}
The applied magnetic field to the Germanene layer causes to the
spin dependent property for one particle Green's function.
In order to perform summation over Matsubara frequency $\omega_{m}$ in Eq.(\ref{e0.702}),
 the Matsubara Green's function elements
should be written in terms of imaginary part of retarded Green's function matrix
 elements using Lehman equation\cite{mahan} as
\begin{eqnarray}
G^{\sigma}_{\alpha\beta}({\bf k},i\omega_{m})=\int^{+\infty}_{-\infty}\frac{d\epsilon}{2\pi}\frac{-2\rm Im
G^{\sigma}_{\alpha\beta}({\bf k},i\omega_{m}\longrightarrow\epsilon+i0^{+})}{i\omega_{m}-\epsilon},
 \label{a0.702}
\end{eqnarray}
where $G^{\sigma}_{\alpha\beta}({\bf k},i\omega_{m}\longrightarrow\epsilon+i0^{+})$ denotes the retarded Green's function
matrix element.
 By replacing Lehman representation for Matsubara Green's function matrix elements
 into Eq.(\ref{e0.702}) and taking summation
over fermionic Matsubara frequency $\omega_{m}$, we obtain dynamical transverse
 spin
susceptibility of electrons on Germanene structure as following form
\begin{eqnarray}
\chi_{+-}({\bf q},i\Omega_{n})&=&-
\frac{1}{N}\sum_{{\bf k}}\sum_{\alpha,\beta}\int^{+\infty}_{-\infty}\int^{+\infty}
_{-\infty}\frac{d\epsilon d\epsilon'}{\pi^{2}}
\rm Im
G^{\uparrow}_{\beta\alpha}({\bf k},\epsilon+i0^{+})\rm Im
G^{\downarrow}_{\alpha\beta}({\bf k}+{\bf q},\epsilon'+i0^{+})\nonumber\\&\times&\frac{n_{F}(\epsilon)-n_{F}(\epsilon')}
{i\Omega_{n}+\epsilon-\epsilon'},
\label{e0.703}
\end{eqnarray}
where $n_{F}(x)=\frac{1}{e^{x/k_{B}T}+1}$ implies well known Fermi-Dirac distortion function.
The Fourier transformation of longitudinal component of the spin, i.e. $S^{z}({\bf q})$, is given in terms of fermionic
operators as
\begin{eqnarray}
 S^{z}({\bf q})=\sum_{{\bf k},\sigma}\sigma\Big(
a^{\dag\sigma}_{{\bf k}+{\bf q}}
a^{\sigma}_{\bf k}+b^{\dag\sigma}_{{\bf k}+{\bf q}}
b^{\sigma}_{\bf k}\Big)
\label{e0.704}
\end{eqnarray}
Also
$\chi_{zz}$ is introduced as longitudinal spin susceptibility and its relation can be expressed in terms of correlation function between
$z$ component of spin operators as
\begin{eqnarray}
\chi_{zz}({\bf q},\omega)&=&i
\int_{-\infty}^{+\infty}dt e^{i\omega t}
\langle [S^{z}({\bf q},t),S^{z}(-{\bf q},0)]\rangle\nonumber\\&=&\lim_{i\Omega_{n}
\longrightarrow\omega+i0^{+}}\int_{0}^{1/(k_{B}T)}d\tau e^{i\Omega_{n}\tau}\langle
\mathcal{T}S^{z}({\bf q},\tau)S^{z}(-{\bf q},0)\rangle\nonumber\\&=&
\chi_{zz}({\bf q},i\Omega_{n}\longrightarrow\omega+i0^{+}).
 \label{a5.6}
\end{eqnarray}
After some algebraic calculations similar to transverse spin
susceptibility case,
 we arrive te final results for Matsubara representation of longitudinal
dynamical spin susceptibility as
\begin{eqnarray}
\chi_{zz}({\bf q},i\Omega_{n})&=&-
\frac{1}{N}\sum_{{\bf k}}\sum_{\alpha,\beta,\sigma}\int^{+\infty}_{-\infty}
\int^{+\infty}_{-\infty}\frac{d\epsilon d\epsilon'}{\pi^{2}}
\rm Im
G^{\sigma}_{\beta\alpha}({\bf k},\epsilon+i0^{+})\rm Im
G^{\sigma}_{\alpha\beta}({\bf k}+{\bf q},\epsilon'+i0^{+})\nonumber\\&\times&\frac{n_{F}(\epsilon)-n_{F}(\epsilon')}
{i\Omega_{n}+\epsilon-\epsilon'},
\label{a0.704}
\end{eqnarray}
The dynamical spin structure factor for both longitudinal and transverse spin
directions are obtained based on retarded presentation of susceptibilities as
\begin{eqnarray}
\chi_{zz}({\bf q},\omega)=\chi_{zz}(q,i\Omega_{n}\longrightarrow\omega+i0^+)\;\;,\;\;
\chi_{+-}({\bf q},\omega)=\chi_{+-}(q,i\Omega_{n}\longrightarrow\omega+i0^+).
 \label{e962.5}
\end{eqnarray}
so that $\chi_{zz}$ and $\chi_{+-}$ are retarded dynamical spin structure factors
for longitudinal and transverse components of spins, respectively.
The imaginary part of retarded dynamical spin structure factor, i.e. $\rm Im\chi_{+-}({\bf q},\omega),\rm Im\chi_{zz}({\bf q},\omega)$, is proportional to
 the contribution of localized spins
in the neutron differential cross-section.
For each ${\bf q}$ the dynamical structure factor has peaks at certain energies which
 represent collective excitations spectrum of the system.
The imaginary parts of both retarded transverse and longitudinal spin structure factors are given by
\begin{eqnarray}
\rm Im\chi_{+-}({\bf q},\omega)&=&
\frac{1}{N}\sum_{{\bf k}}\sum_{\alpha,\beta}\int^{+\infty}_{-\infty}\frac{d\epsilon}{\pi}
\rm Im
G^{\uparrow}_{\beta\alpha}({\bf k},\epsilon+i0^{+})\rm Im
G^{\downarrow}_{\alpha\beta}({\bf k}+{\bf q},\epsilon+\omega+i0^{+})\nonumber\\&\times&\Big(n_{F}(\epsilon)-n_{F}(\epsilon+\omega)\Big)
,\nonumber\\
\rm Im\chi_{zz}({\bf q},\omega)&=&
\frac{1}{N}\sum_{{\bf k}}\sum_{\alpha,\beta,\sigma}\int^{+\infty}_{-\infty}\frac{d\epsilon}{\pi}
\rm Im
G^{\sigma}_{\beta\alpha}({\bf k},\epsilon+i0^{+})\rm Im
G^{\sigma}_{\alpha\beta}({\bf k}+{\bf q},\epsilon+\omega+i0^{+})\nonumber\\&\times&\Big(n_{F}(\epsilon)-n_{F}(\epsilon+\omega)\Big)
.
\label{e0.963}
\end{eqnarray}
The frequencies of collective magnetic excitation modes are determined via finding the
position of peaks in imaginary part of of
imaginary part of dynamical
spin susceptibilities.

Static transverse spin
structure factor ($s_{+-}({\bf q})$) which is a measure of
magnetic long range ordering for spin components along the plane, i.e. transverse direction,
 can be related to imaginary part of
retarded dynamical spin susceptibility using following relation
\begin{eqnarray}
s_{+-}({\bf q})=
\left\langle S^{+}({\bf q})S^{-}(-{\bf q}) \right\rangle&=&
k_{B}T\sum_{n}\frac{1}{2\pi}
\int_{-\infty}^{\infty}d\omega\frac{-2\rm Im\chi_{+-}({\bf q},i\omega_{n}\longrightarrow\omega+i0^+)}{i\omega_{n}-\omega}\nonumber\\
&=&\int_{-\infty}^{+\infty}d\omega\frac{n_{B}(\omega)}{\pi}\rm Im\chi_{+-}
({\bf q},i\omega_{n}\longrightarrow\omega+i0^+).
 \label{e9631}
\end{eqnarray}
Moreover can find the static longitudinal spin structure factor, i.e. $s_{+-}({\bf q})$, by using $\rm Im\chi_{zz}({\bf q},\omega)$ as
\begin{eqnarray}
s_{zz}({\bf q})&=&
\left\langle S^{z}({\bf q})S^{z}(-{\bf q}) \right\rangle=
k_{B}T\sum_{n}\frac{1}{2\pi}
\int_{-\infty}^{\infty}d\omega\frac{-2\rm Im\chi_{zz}({\bf q},i\omega_{n}\longrightarrow\omega+i0^+)}{i\omega_{n}-\omega}\nonumber\\
&=&\int_{-\infty}^{+\infty}d\omega\frac{n_{B}(\omega)}{\pi}\rm Im\chi_{zz}
({\bf q},i\omega_{n}\longrightarrow\omega+i0^+).
 \label{e9632}
\end{eqnarray}
In the next section, the numerical results of dynamical spin structure and static spin
structures of Germanene layer have been presented for various magnetic field and spin-orbit coupling strength.
\section{Numerical Results and Discussions}
We turn to a presentation of our main numerical results of imaginary
part of dynamical structure factors of Germanene layer at
finite temperature in the presence of magnetic field and spin-orbit coupling.
Also the temperature dependence of static structure factors has been
addressed in this section.
Using the electronic band structure, the matrix elements of Fourier transformations of spin dependent Green's function
 are calculated according to Eq.(\ref{e5.5}). The imaginary part of both
 transverse and longitudinal dynamical spin susceptibilities is made by substituting the Green's function
matrix elements into Eq.(\ref{e0.963}).
In the following, the frequency behavior of imaginary part of dynamical spin
susceptibilities is studied at fixed wave number
${\bf q}=(0,\frac{4\pi}{3})$ in the Brillouin zone where the length of unit cell vector of honeycomb
 lattice is taken to be unit.
Furthermore the static structure factors have been obtained by using Eqs.(\ref{e9631},\ref{e9632}).

The frequency behaviors of both the transverse and longitudinal
 dynamical spin susceptibilities have been addressed in this present study.
Also the spin structure factors behaviors have been investigated for Germanene structure.

The optimized atomic structure of the Germanene with primitive unit cell vector length $a=1$
is shown in Fig.(1). The primitive unit cell include two Ge atoms.

 In Fig.(3), we depict the frequency dependence of imaginary part of longitudinal
 dynamical spin susceptibility, $Im\chi_{zz}({\bf q}_{0},\omega)$, of undoped Germanene layer
 for different values of spin-orbit coupling, namely $\lambda/t=0.08,0.12,0.16,0.2$,
in the absence of magnetic field by setting $k_{B}T/t=0.05$.
 In fact the effects of spin-orbit coupling strength on
frequency dependence
of $Im\chi_{zz}({\bf q},\omega)$ have been studied in this figure.
 As shown in
Fig.(3), the frequency positions of sharp peaks in  $Im\chi_{zz}({\bf q}_{0},\omega)$,
that imply
 spin excitation mode for longitudinal components of spins,
 moves to higher frequencies with increase of spin-orbit coupling.
This fact can be understood from this point that the increase of spin-orbit coupling
leads to enhance band gap in density of
states and consequently the excitation mode appears in higher frequency.
 Note that this figure shows the
inelastic cross section neutron particles from itinerant electrons of the system due
 to longitudinal component along $z$ direction of
magnetic moment of electrons and neutron beam.
Another novel feature in Fig.(3) is the increase of intensity of sharp peak with $\lambda$.
For frequencies above normalized value 1.5, there is no collective magnetic excitation mode
for longitudinal components of
electron spins as shown in Fig.(3).

The frequency dependence of imaginary part of longitudinal dynamical spin
structure factor of undoped Germanene layer in the presence of spin polarization
 for different magnetic field values $g\mu_{B}B/t$ has been shown
in Fig.(4) at fixed $\lambda/t=0.6$ by setting $k_{B}T/t=0.05$.
 A novel feature has been pronounced in this figure. It is clearly observed that all curves for different magnetic field indicates
two clear magnetic excitation collective modes at frequencies $\omega/t\approx0.45$ and $\omega/t\approx2.1$.
The frequency positions of magnetic excitation mode is independent of magnetic field.
The intensity of sharp peaks in $\omega/t\approx2.1$ in imaginary part of longitudinal susceptibility, i.e. $Im\chi_{zz}({\bf q}_{0},\omega)$,
 decreases with magnetic field.
However the height of sharp peaks in frequency position $\omega/t\approx0.45$ enhances with increase
of spin-orbit coupling as shown in Fig.(4).

In Fig.(5), the imaginary part of longitudinal dynamical spin structure factor of Germanene layer
 has been plotted for different values of chemical potential, namely $\mu/t=0.2,0.3,0.4,0.5$,
at fixed spin-orbit coupling $\lambda/t=0.3$ by setting $k_{B}T/t=0.05$ in the absence of magnetic field.
This figure implies the frequency position and intensity of collective excitation mode in $\omega/t\approx1.5$
has no dependence on chemical potential. Although the intensity of sharp peak in $Im\chi_{zz}({\bf q}_{0},\omega)$ at frequency
position $\omega/t\approx0.4$ increases with chemical potential. There is no considerable change for frequency position in
$Im\chi_{zz}({\bf q}_{0},\omega)$ at $\omega/t\approx0.4$ with chemical potential according to Fig.(5).
Also the intensity of low energy magnetic excitation mode reduces with decreasing chemical potential value $\mu$.

The behavior of longitudinal static spin structure
factor $s_{zz}({\bf q}_{0})$ of undoped Germanene layer in terms of normalized temperature
$k_{B}T/t$ for different values of $\lambda/t$
has been presented in Fig.(6). The applied magnetic field is assumed to be zero.
This function is a measure for the tendency to magnetic long range ordering for the longitudinal components of spins in the
itinerant electrons.
This figure implies static spin structure for spin components perpendicular to the Germanene plane includes a peak for each value of $\lambda/t$.
The temperature position of peak in longitudinal spin static structure factor is the same for all spin-orbit coupling
strengths. Although the height of peak increases with $\lambda/t$ which justifies the long range ordering for $z$ components of spins
improves with spin-orbit coupling. Another novel feature is the non zero value for static structure factor
$s_{zz}({\bf q}_{0})$ at zero temperature. However the increase of temperature up to peak position
 leads to raise magnetic long range ordering. Upon more increasing temperature above normalized value 0.25, the thermal fluctuations
causes to reduce $s_{zz}({\bf q}_{0})$ so that magnetic long range ordering of the electrons decays as shown in Fig.(6).

The temperature dependence of static longitudinal spin structure factor of doped Germanene layer for various
chemical potential has been studied in Fig.(7) for $g\mu_{B}B/t=0.0$ by setting $\lambda/t=0.4$.
In contrast to the undoped case in Fig.(6), it is clearly observed the longitudinal spin structure factor for each value
finite chemical potential gets zero value at zero temperature
according to Fig.(7). In fact the quantum fluctuations at zero temperature
for doped case leads to destroy any magnetic long range ordering.
Moreover the peak in temperature dependence of $s_{zz}({\bf q}_{0})$ tends to higher temperature upon increasing electron doping.
Also the height of peak ,as a measure of magnetic long range ordering for longitudinal components of spins,
reduces with chemical potential according to Fig.(7).
Upon increasing temperature above normalized value 1.5,  $s_{zz}({\bf q}_{0})$ increases with $\mu/t$ and consequently the increase of
electron doping improves the long range ordering in temperature region above normalized value 1.5.

The effect of longitudinal magnetic field on the behavior of $s_{zz}({\bf q}_{0})$ in terms of normalized temperature
$k_{B}T/t$ in undoped case for different values of magnetic field, namely $g\mu_{B}B/t=0.3,0.4,0.5,0.6,0.7$ has
 been plotted in Fig.(8). The static structure is considerably affected by magnetic field at low temperatures below normalized amount
0.5 where the quantum effects are more remarkable. In addition, at fixed values of temperatures above normalized value 0.5, $s_{zz}({\bf q}_{0})$
is independent of magnetic field and all curves fall on each other on the whole range of temperature in this temperature region.
Also temperature position of peak in longitudinal static spin structure factor moves to lower temperature with increasing magnetic field
according to Fig.(8). Moreover the height of peak enhances with magnetic field. It can be understood from the fact that applying magnetic
field along $z$ direction perpendicular to the plane causes long range ordering of $z$ components of spins of electrons which increases
the longitudinal static structure factor at low temperatures below 0.5.

We have also studied the dynamical and static transverse spin structure factors of Germanene layer in the presence of
magnetic field and spin-orbit coupling.
Our main results for imaginary part of dynamical transverse spin susceptibility of undoped Germanene layer
 for different spin-orbit coupling strengths at fixed temperature $k_{B}T/t=0.05$ in the absence of magnetic field are
summarized in
 Fig.(9). The collective magnetic excitation modes for spin components parallel to the plane
 tends to higher values with $\lambda/t$ according to Fig.(9).
This feature arises from the increase of band gap with spin-orbit coupling so that collective mode appears at higher frequencies as
shown in Fig.(9).
Also the height of peak in $\rm Im\chi_{+-}({\bf q}_{0},\omega)$, which is proportional to intensity of
scattered neutron beam from itinerant electrons of Germanene structure, reduces with decreasing spin-orbit coupling strength.

In Fig.(10) we plot the numerical results of $\rm Im\chi_{+-}({\bf q},\omega)$ of undoped Germanene layer
 as a function
of normalized frequency $\omega/t$ for
various magnetic field, namely $g\mu_{B}B/t=0.0,0.2,0.4,0.6$ by setting $k_{B}T/t=0.05$.
It is clearly observed that the frequency position of collective magnetic excitation mode moves to lower values with magnetic field.
It can be understood from the fact that the applying magnetic field gives rise to reduce the band gap and consequently the
excitation mode takes place at lower frequencies.
Moreover the intensity of collective excitation mode for transverse components of spins is clearly independent of
magnetic field strength according to Fig.(10).

 Fig.(11) presents the effect of electron doping on the
frequency dependence of $\rm Im\chi_{+-}({\bf q}_{0},\omega)$ of Germanene layer
by setting $\lambda/t=0.3$ at fixed value
of temperature in the absence of magnetic field.
There are two collective magnetic excitation mode for each value chemical potential.
The frequency positions of excitation mode are the same for all values of chemical potential.
The intensity of low frequency peak increases with chemical potential however the intensity of
 high frequency peak reduces with electron doping as shown in Fig.(11).
The increase of electron doping leads to decrease transition rate of electron from valence band to conduction one and consequently
the intensity of excitation mode decreases.

The behavior of transverse static spin structure
factor $s_{+-}({\bf q}_{0})$ of undoped Germanene layer as a function of normalized temperature
$k_{B}T/t$ for different values of $\lambda/t$ in the absence of magnetic field
has been presented in Fig.(12).
This function is a measure for the tendency to magnetic long range ordering for the transverse components of spins in the
itinerant electrons.
A peak in $s_{+-}({\bf q}_{0})$ is clearly observed for each value of $\lambda/t$.
The peak is located at 0.25 for all values of spin-orbit coupling
strengths. Although the height of peak increases with $\lambda/t$ which justifies the long range ordering for transverse components of spins
improves with spin-orbit coupling. Another novel feature is the non zero value for static structure factor
$s_{+-}({\bf q}_{0})$ at zero temperature. In temperature region below peak position, the increase of temperature
 leads to raise magnetic long range ordering. Upon more increasing temperature above normalized value 0.25, the thermal fluctuations
causes to reduce $s_{zz}({\bf q}_{0})$ and magnetic long range ordering of the electrons as shown in fig.(12).

The temperature dependence of static transverse spin structure factor of doped Germanene layer for various
chemical potential in the absence of magnetic field has been studied in Fig.(13) by setting $\lambda/t=0.3$.
In contrast to the undoped case in Fig.(12), it is clearly observed the transverse spin structure factor for each value
finite chemical potential gets zero value at zero temperature
according to Fig.(13). In fact the quantum fluctuations at zero temperature
for doped case leads to destroy any magnetic long range ordering.
Moreover the temperature position of peak in $s_{+-}({\bf q}_{0})$ appears in $k_{B}T/t=0.5$ for all amounts of chemical potential.
Also the height of peak ,as a measure of magnetic long range ordering for transverse components of spins,
reduces with chemical potential according to Fig.(13). In other words the increase of electron doping leads to decrease magnetic long range ordering
for transverse components of spins.

The effect of longitudinal magnetic field on the behavior of $s_{+-}({\bf q}_{0})$ in terms of normalized temperature
$k_{B}T/t$ in undoped case for different values of magnetic field, namely $g\mu_{B}B/t=0.4,0.45,0.5,0.55,0.6$ has
 been plotted in Fig.(14). The static transverse spin structure factor is considerably affected by magnetic field at low temperatures
 below normalized amount
0.25 where the quantum effects are more remarkable. In addition, at fixed values of temperatures above normalized value 0.25, $s_{+-}({\bf q}_{0})$
is independent of magnetic field and all curves fall on each other on the whole range of temperature in this temperature region.
Also temperature position of peak in longitudinal static spin structure factor moves to lower temperature with increasing magnetic field.
Moreover the height of peak enhances with magnetic field. It can be understood from the fact that applying magnetic
field along $z$ direction perpendicular to the plane causes long range ordering of components of spins parallel to the plane
 of electrons which increases
the longitudinal static structure factor at low temperatures below 0.25.

\section{Conflict Of interest Statement}
There is no conflict of interest statement in this manuscript.

\begin{figure}
\begin{center}
\epsfxsize=0.8\textwidth
\includegraphics[width=12.cm]{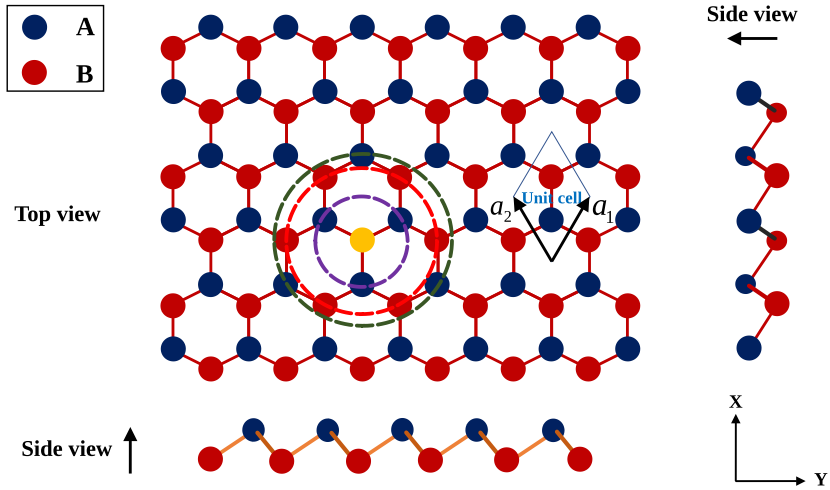}
\small
\begin{flushleft}
\caption{Crystal structure of Germanene}
\end{flushleft}
\end{center}
\end{figure}
\begin{figure}
\begin{center}
\epsfxsize=0.8\textwidth
\includegraphics[width=12.cm]{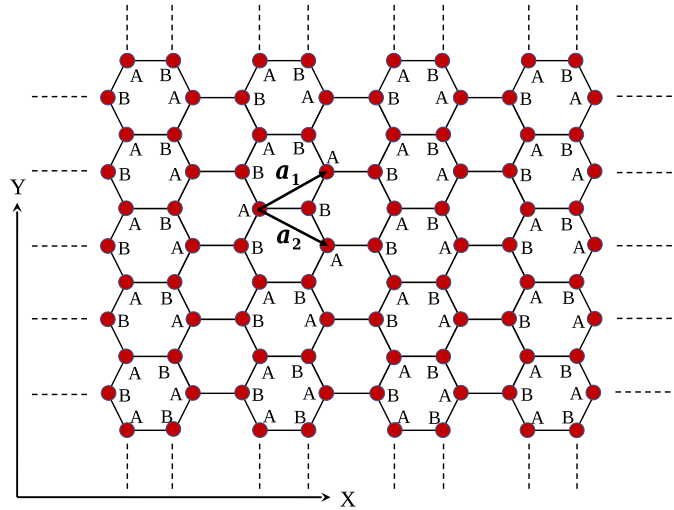}
\small
\begin{flushleft}
\caption{The structure of honeycomb structure is shown. The
light dashed lines denote the Bravais lattice unit cell. Each cell
includes two nonequivalent sites, which are indicated by A and B.
${\bf a}_{1}$ and ${\bf a}_{2}$ are the primitive vectors of unit
cell.}
\end{flushleft}
\end{center}
\end{figure}
\begin{figure}
\begin{center}
\epsfxsize=0.8\textwidth
\includegraphics[width=12.cm]{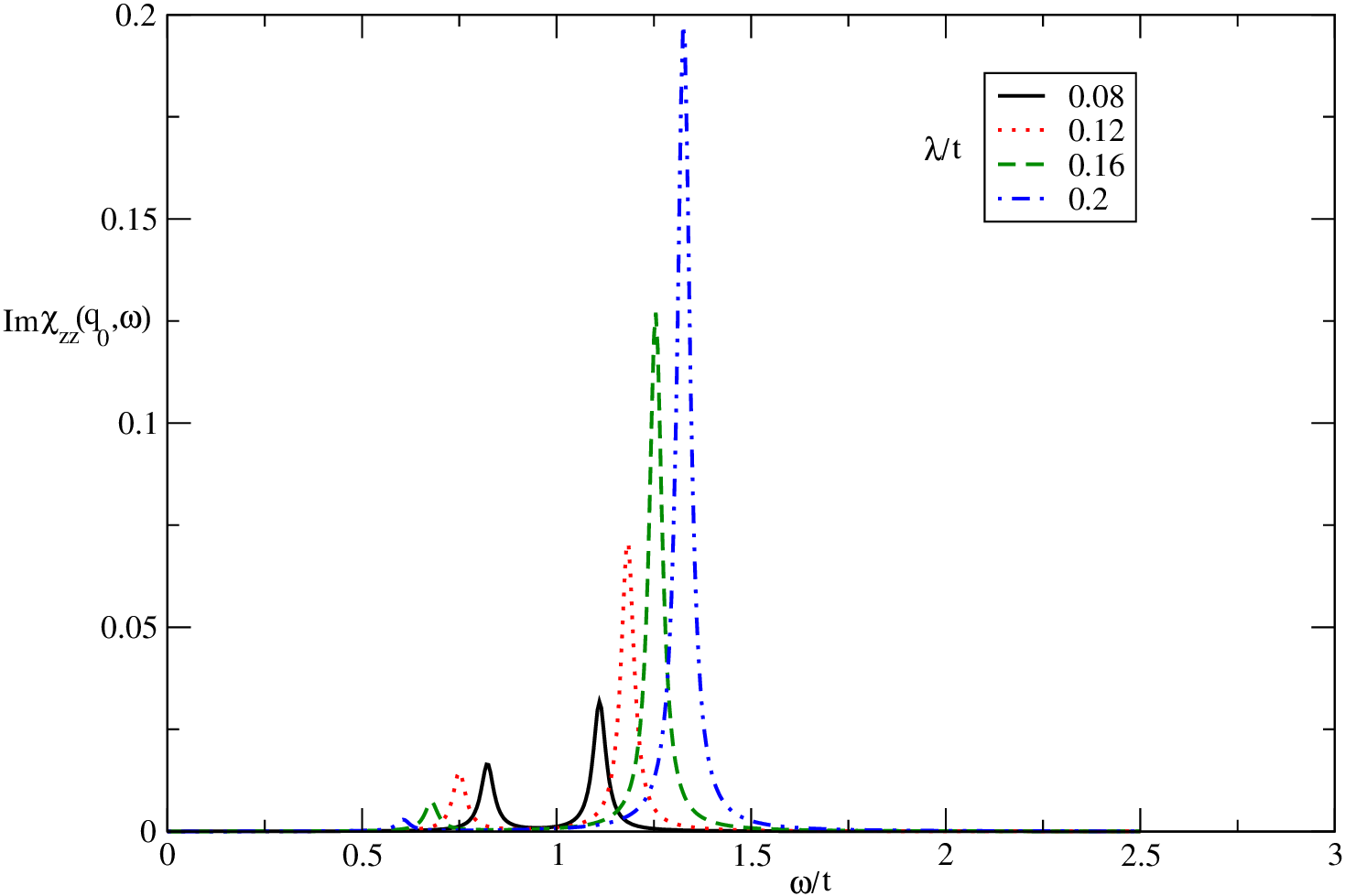}
\small
\begin{flushleft}
\caption{The imaginary part of dynamical longitudinal
 spin susceptibility $\rm Im\chi_{zz}({\bf q}_{0},\omega)$ of
 undoped Germanene layer versus normalized frequency $\omega/t$
for different values of normalized spin-orbit coupling
$\lambda/t$ at fixed temperature $k_{B}T/t=0.05$.
The magnetic field is considered to be zero.}
\end{flushleft}
\end{center}
\end{figure}

\begin{figure}
\begin{center}
\epsfxsize=0.8\textwidth
\includegraphics[width=12.cm]{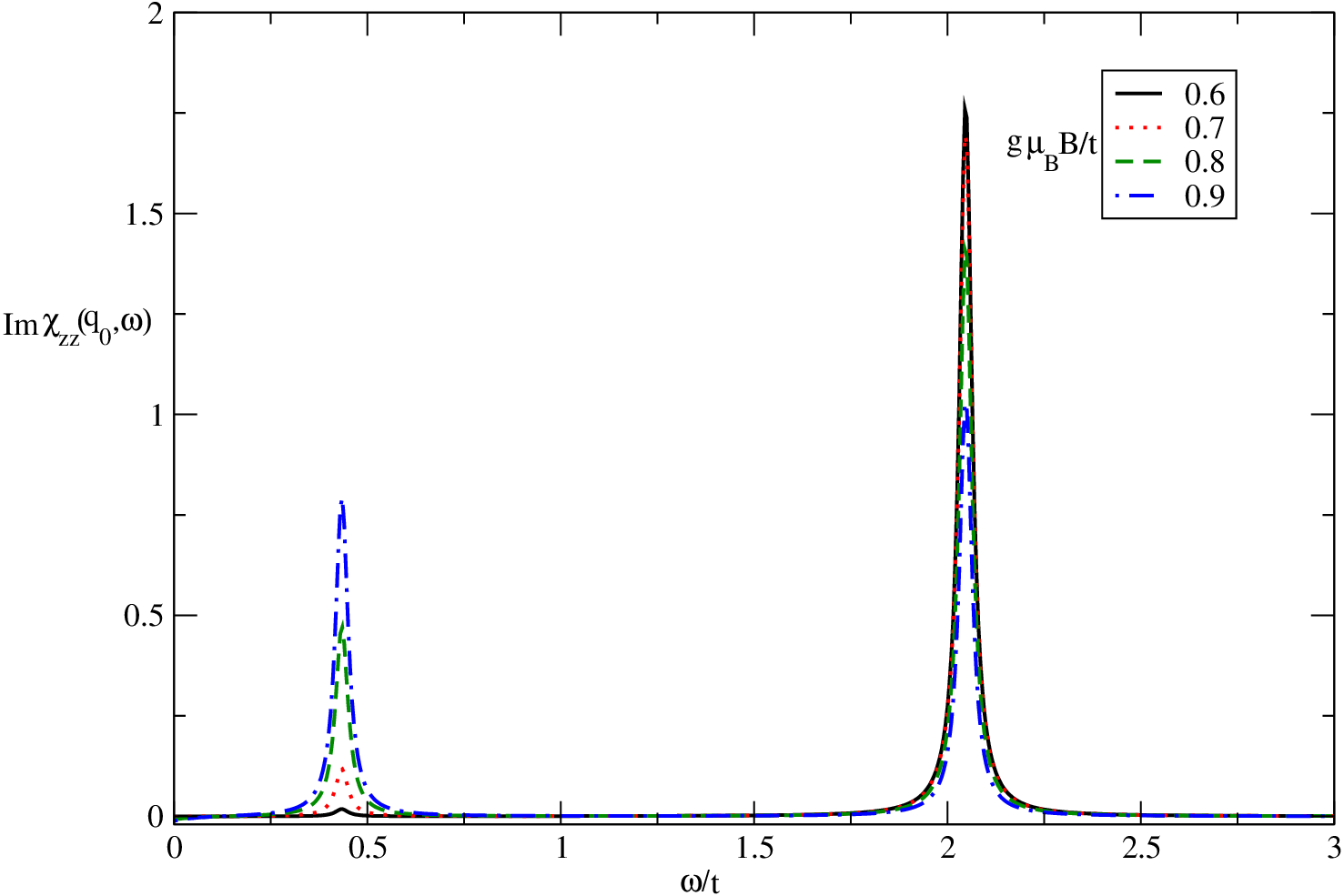}
\small
\begin{flushleft}
\caption{The imaginary part of dynamical longitudinal
 spin susceptibility $\rm Im\chi_{zz}({\bf q}_{0},\omega)$ of
 undoped Germanene layer versus normalized frequency $\omega/t$
for different values of normalized magnetic field
$g\mu_{B}B/t$ at fixed spin-orbit coupling $\lambda/t=0.6$.
The normalized temperature is considered to be $k_{B}T/t=0.05$.}
\end{flushleft}
\end{center}
\end{figure}

\begin{figure}
\begin{center}
\epsfxsize=0.8\textwidth
\includegraphics[width=12.cm]{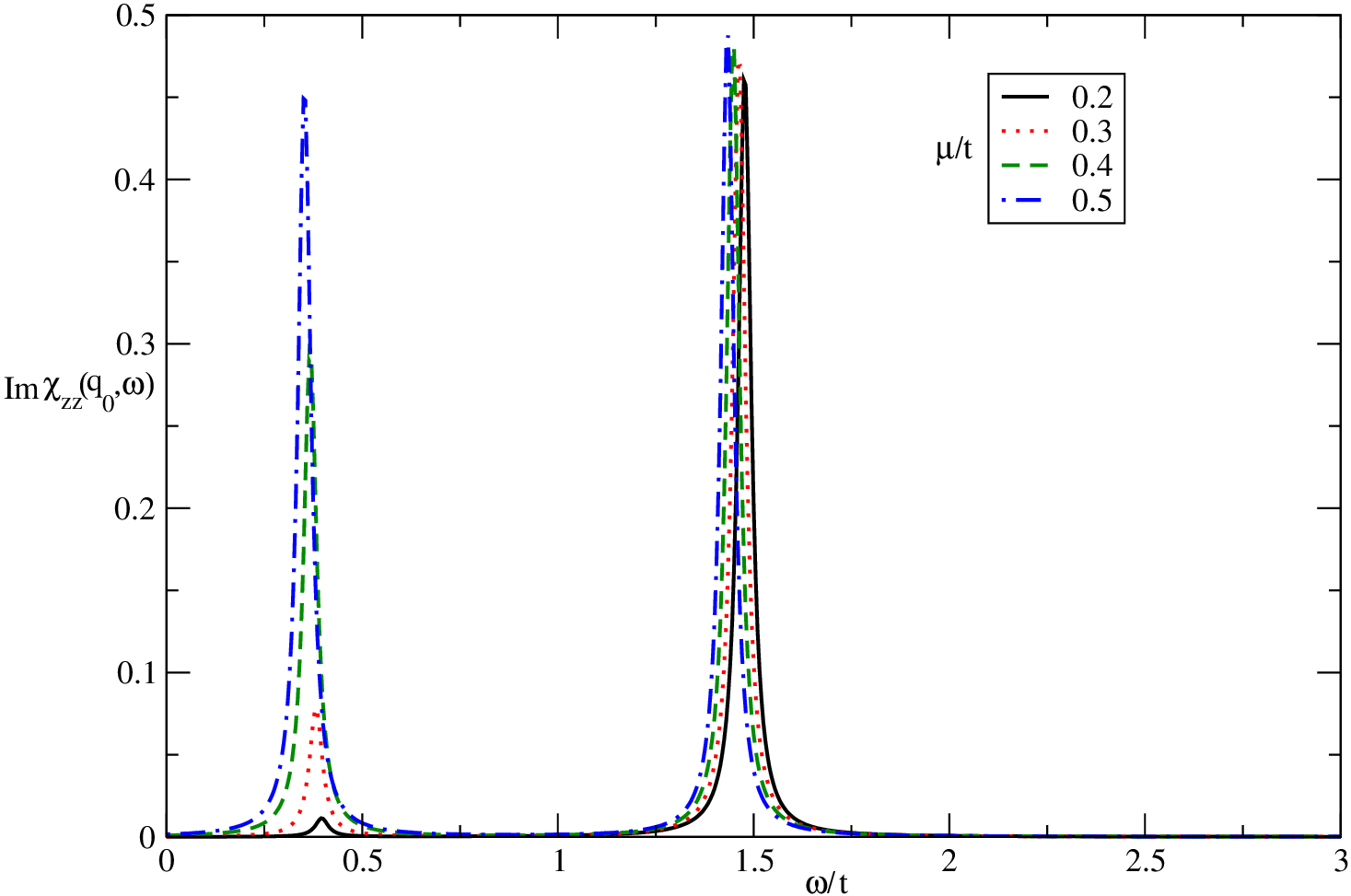}
\small
\begin{flushleft}
\caption{The imaginary part of dynamical longitudinal
 spin susceptibility $\rm Im\chi_{zz}({\bf q}_{0},\omega)$ of
 doped Germanene layer versus normalized frequency $\omega/t$
for different values of normalized chemical potential
$\mu/t$ at fixed spin-orbit coupling $\lambda/t=0.3$.
The magnetic field is assumed to be zero.
The normalized temperature is considered to be $k_{B}T/t=0.05$.}
\end{flushleft}
\end{center}
\end{figure}

\begin{figure}
\begin{center}
\epsfxsize=0.8\textwidth
\includegraphics[width=12.cm]{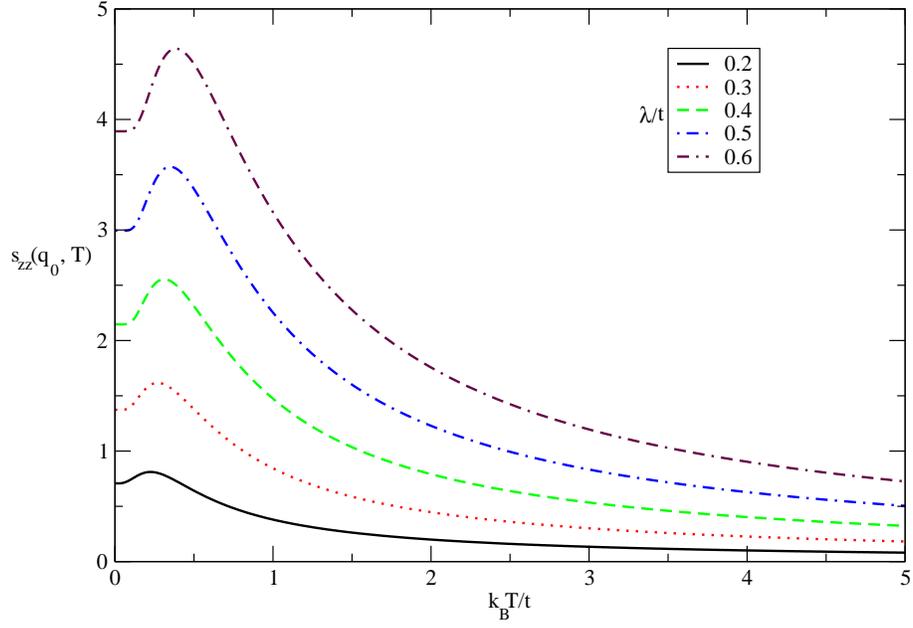}
\small
\begin{flushleft}
\caption{The longitudinal static
 spin structure factor $s_{zz}({\bf q}_{0},T)$ of
 undoped Germanene layer versus normalized temperature $k_{B}T/t$
for different values of spin-orbit coupling strength $\lambda/t$
in the absence of magnetic field.}
\end{flushleft}
\end{center}
\end{figure}
\begin{figure}
\begin{center}
\epsfxsize=0.8\textwidth
\includegraphics[width=12.cm]{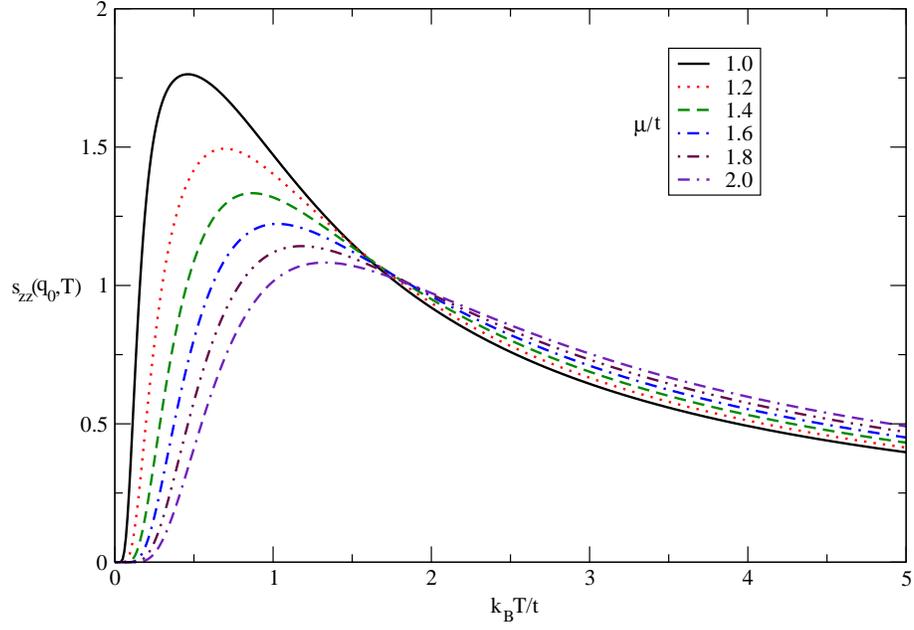}
\small
\begin{flushleft}
\caption{The longitudinal static
 spin structure factor $s_{zz}({\bf q}_{0},T)$of
 undoped Germanene layer versus normalized temperature $k_{B}T/t$
for different values of chemical potential $\mu/t$
in the absence of magnetic field.
Spin-orbit coupling strength has been fixed at $\lambda/t=0.4$.}
\end{flushleft}
\end{center}
\end{figure}

\begin{figure}
\begin{center}
\epsfxsize=0.8\textwidth
\includegraphics[width=12.cm]{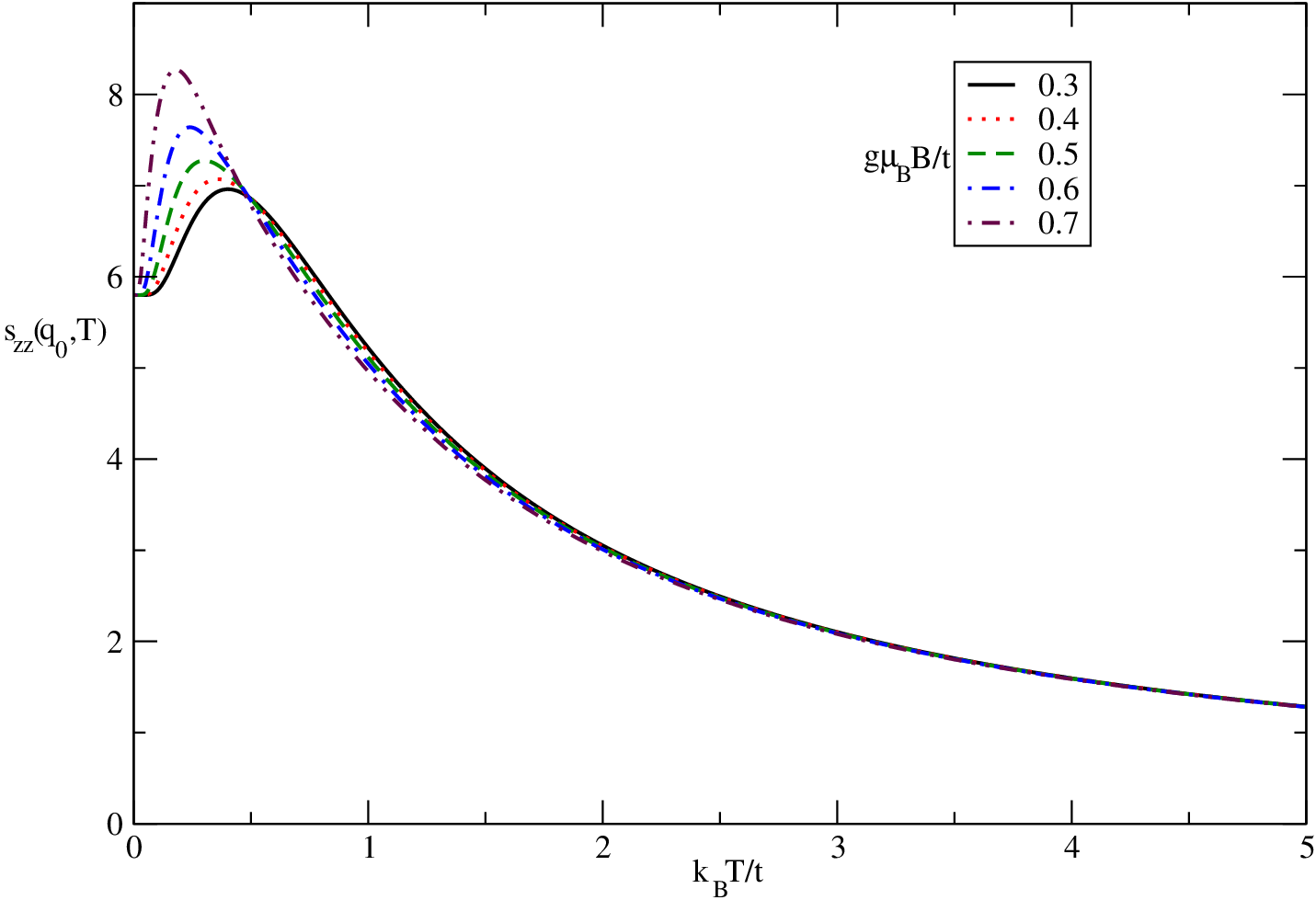}
\small
\begin{flushleft}
\caption{The longitudinal static
 spin structure factor $s_{zz}({\bf q}_{0},T)$ of
 undoped Germanene layer versus normalized temperature $k_{B}T/t$
for different values of magnetic field $g\mu_{B}B/t$
at fixed spin-orbit coupling strength $\lambda/t=0.4$.}
\end{flushleft}
\end{center}
\end{figure}

\begin{figure}
\begin{center}
\epsfxsize=0.8\textwidth
\includegraphics[width=12.cm]{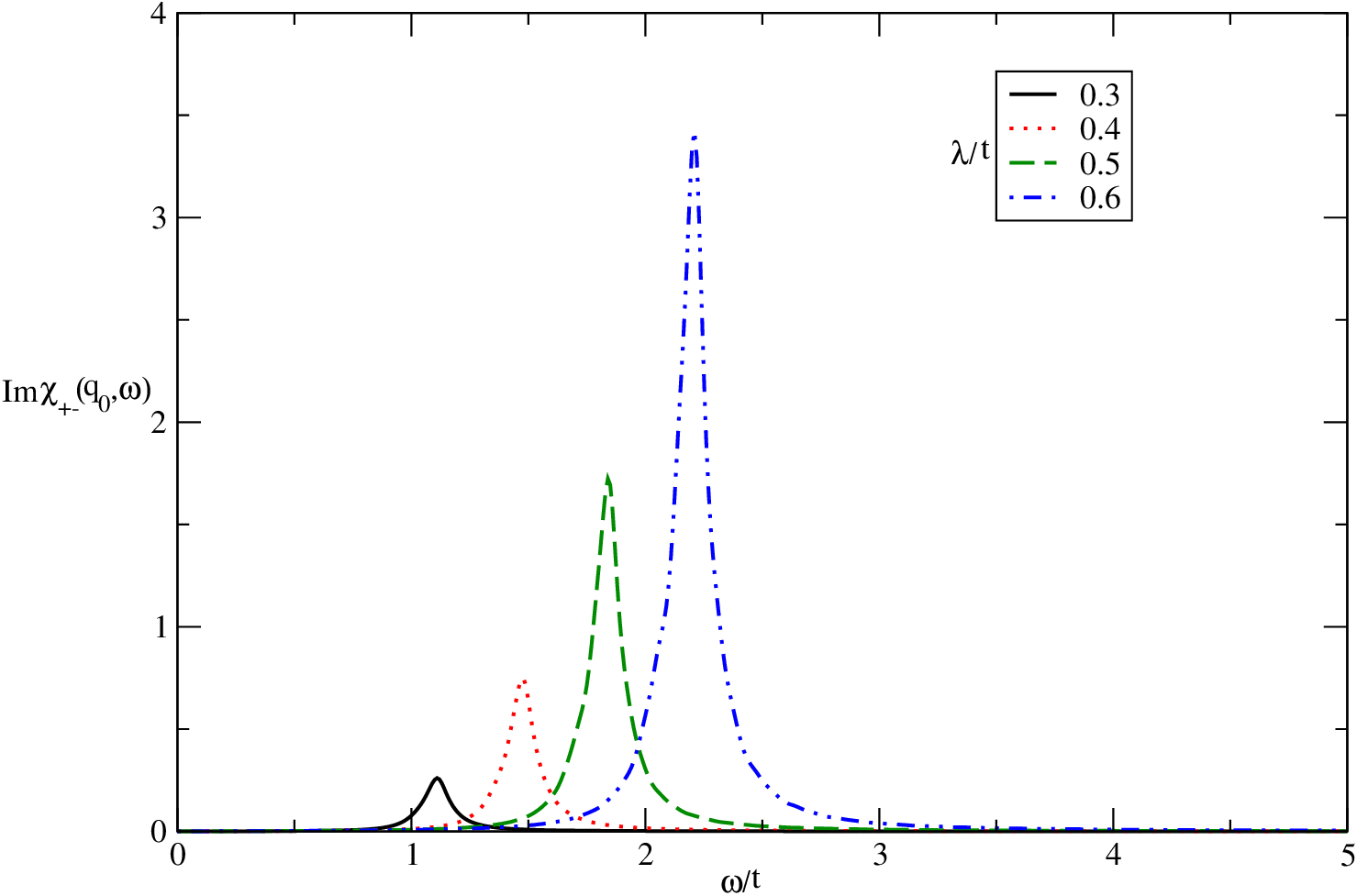}
\small
\begin{flushleft}
\caption{The imaginary part of dynamical transverse
 spin susceptibility $\rm Im\chi_{+-}({\bf q}_{0},\omega)$ of
 undoped Germanene layer versus normalized frequency $\omega/t$
for different values of normalized spin-orbit coupling
$\lambda/t$ at fixed temperature $k_{B}T/t=0.05$.
The magnetic field is considered to be zero. }
\end{flushleft}
\end{center}
\end{figure}

\begin{figure}
\begin{center}
\epsfxsize=0.8\textwidth
\includegraphics[width=12.cm]{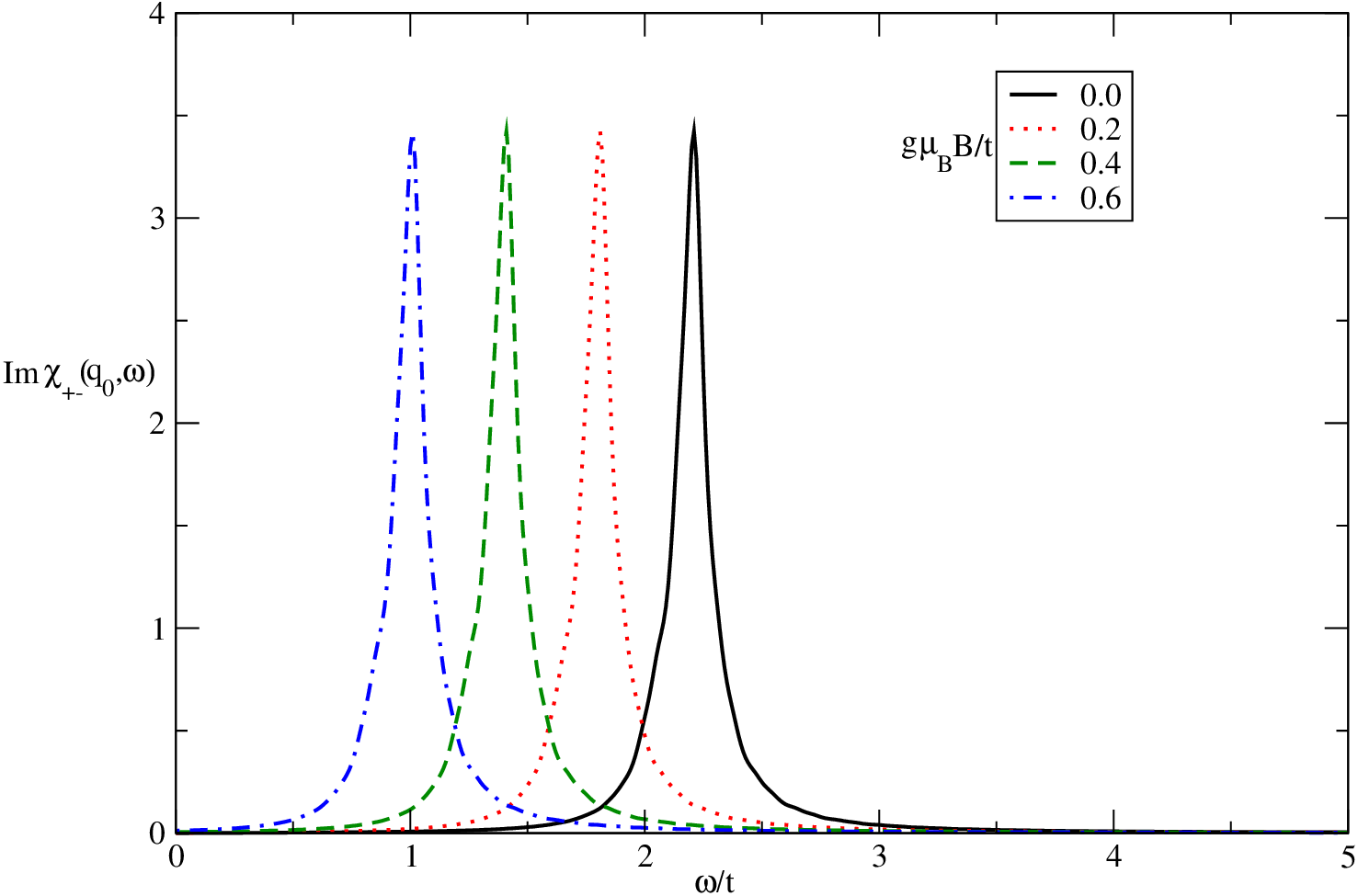}
\small
\begin{flushleft}
\caption{The imaginary part of dynamical transverse
 spin susceptibility $\rm Im\chi_{+-}({\bf q}_{0},\omega)$ of
 undoped Germanene layer versus normalized frequency $\omega/t$
for different values of normalized magnetic field
$g\mu_{B}B/t$ at fixed spin-orbit coupling $\lambda/t=0.6$.
The normalized temperature is considered to be $k_{B}T/t=0.05$.}
\end{flushleft}
\end{center}
\end{figure}

\begin{figure}
\begin{center}
\epsfxsize=0.8\textwidth
\includegraphics[width=12.cm]{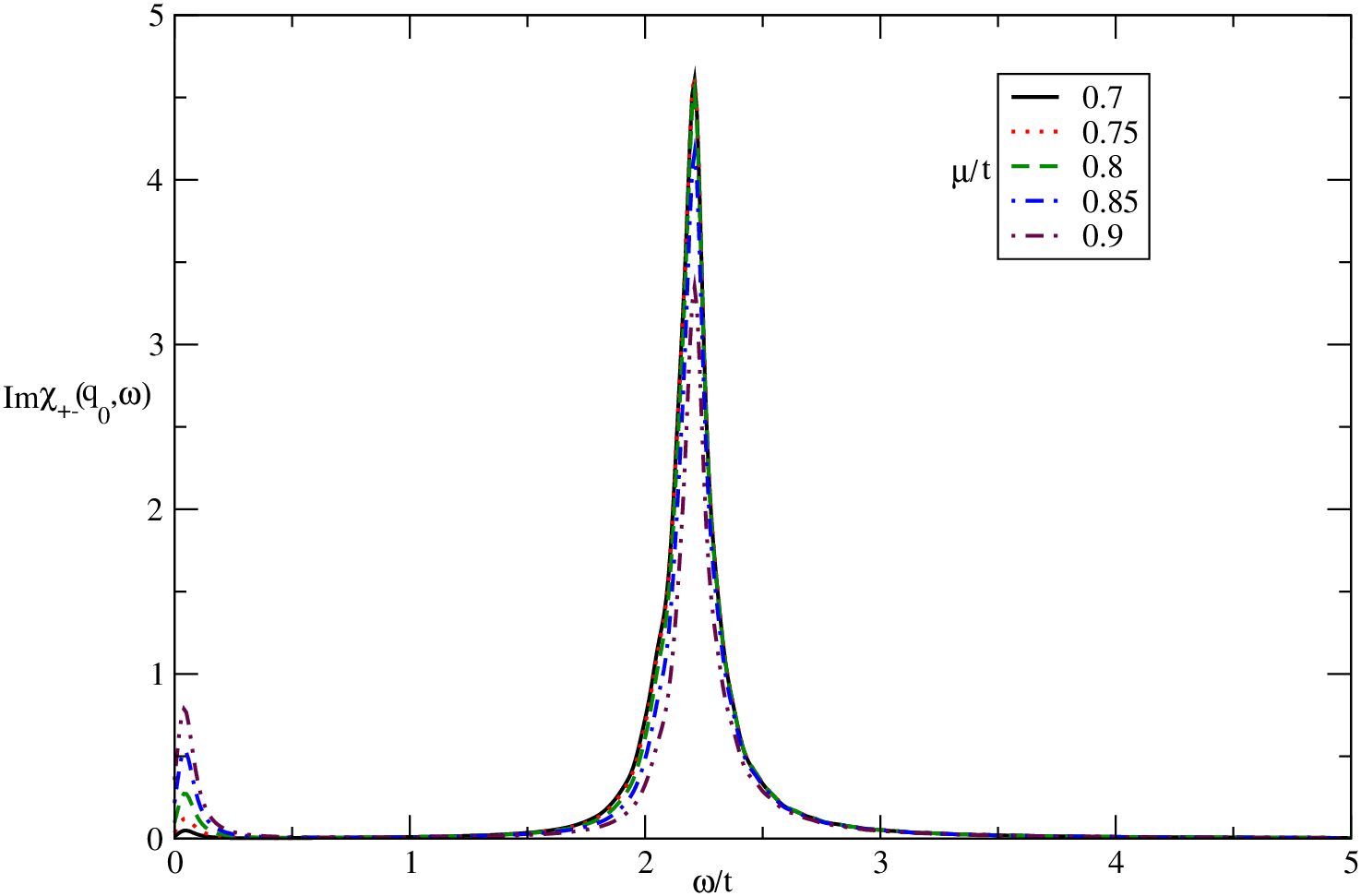}
\small
\begin{flushleft}
\caption{The imaginary part of dynamical transverse
 spin susceptibility $\rm Im\chi_{+-}({\bf q}_{0},\omega)$ of
 doped Germanene layer versus normalized frequency $\omega/t$
for different values of normalized chemical potential
$\mu/t$ at fixed spin-orbit coupling $\lambda/t=0.3$.
The magnetic field is assumed to be zero.
The normalized temperature is considered to be $k_{B}T/t=0.05$. }
\end{flushleft}
\end{center}
\end{figure}
\begin{figure}
\begin{center}
\epsfxsize=0.8\textwidth
\includegraphics[width=12.cm]{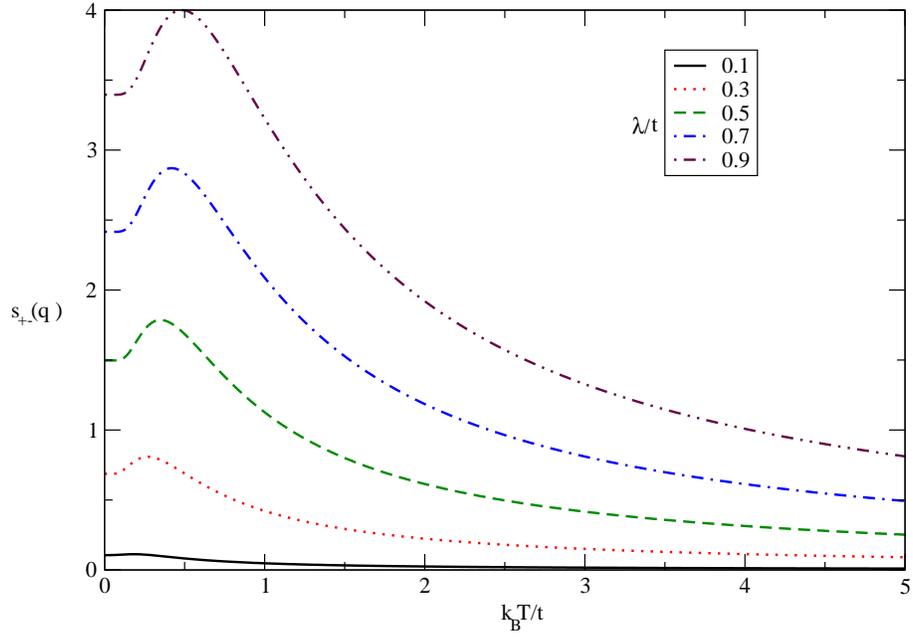}
\small
\begin{flushleft}
\caption{The transverse static
 spin structure factor $s_{+-}({\bf q}_{0},T)$ of
 undoped Germanene layer versus normalized temperature $k_{B}T/t$
for different values of spin-orbit coupling strength $\lambda/t$
in the absence of magnetic field. }
\end{flushleft}
\end{center}
\end{figure}

\begin{figure}
\begin{center}
\epsfxsize=0.8\textwidth
\includegraphics[width=12.cm]{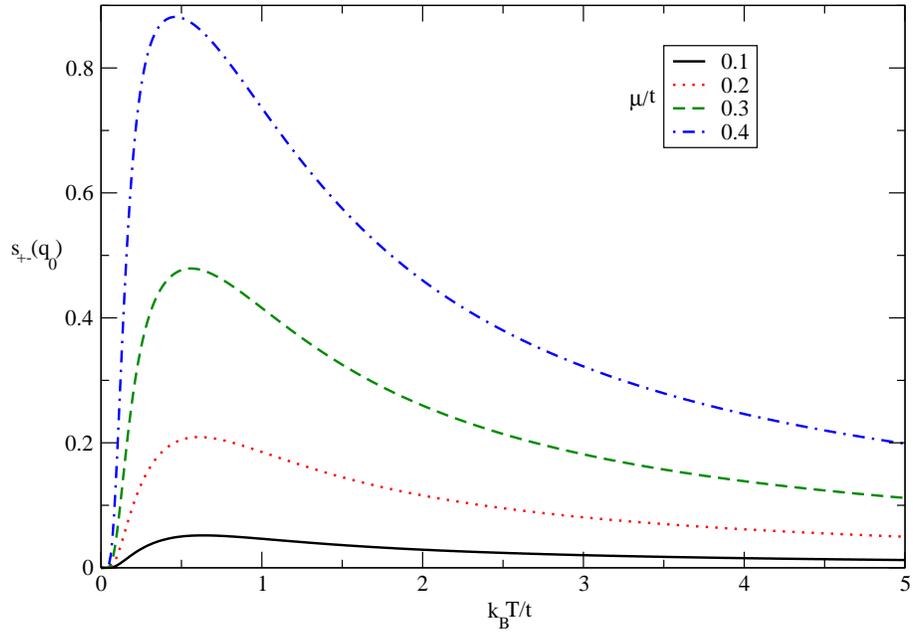}
\small
\begin{flushleft}
\caption{The transverse static
 spin structure factor $s_{+-}({\bf q}_{0},T)$ of
 doped Germanene layer versus normalized temperature $k_{B}T/t$
for different values of chemical potential$\mu/t$
in the absence of magnetic field with spin-orbit coupling $\lambda/t=0.3$.}
\end{flushleft}
\end{center}
\end{figure}

\begin{figure}
\begin{center}
\epsfxsize=0.8\textwidth
\includegraphics[width=12.cm]{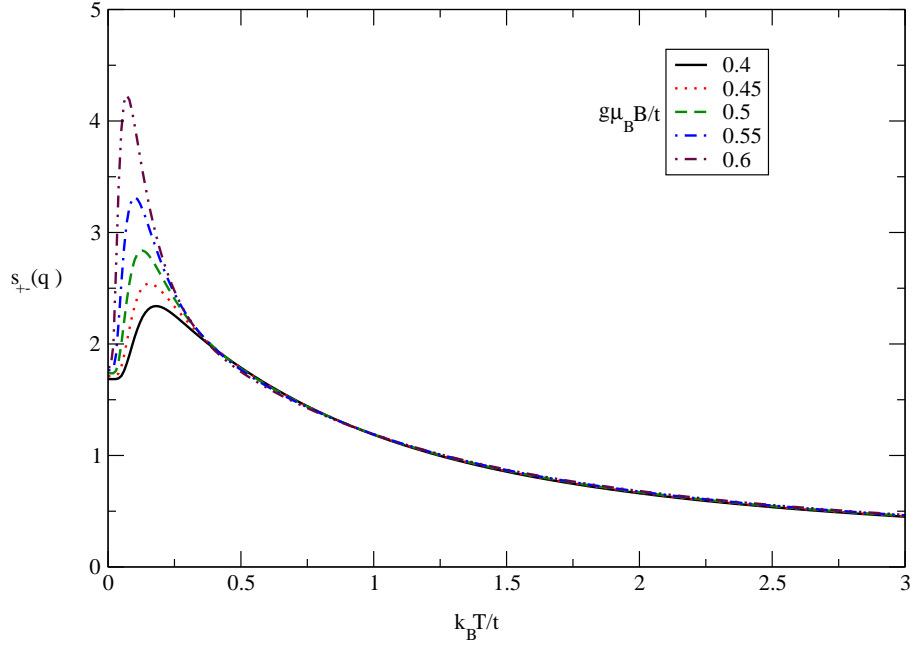}
\small
\begin{flushleft}
\caption{The transverse static
 spin structure factor $s_{+-}({\bf q}_{0},T)$of
 undoped Germanene layer versus normalized temperature $k_{B}T/t$
for different values of magnetic field $g\mu_{B}B/t$.
Spin-orbit coupling strength has been fixed at $\lambda/t=0.4$.}
\end{flushleft}
\end{center}
\end{figure}

\newpage

\end{document}